\documentclass[prl,twocolumn,groupedaddress,showpacs,floatfix]{revtex4-1}
\usepackage{amsmath,amssymb,amsfonts}
\usepackage{mathcomp}
\usepackage{comment}
\usepackage{graphicx,subfigure}
\usepackage{txfonts}
\usepackage{epstopdf}
\usepackage[title]{appendix}

\usepackage[usenames, dvipsnames]{color}
\usepackage{soul}
\usepackage[normalem]{ulem}

\usepackage{bbold}
\usepackage{footnote}

\usepackage{xr-hyper} 
\usepackage[unicode]{hyperref} 
\hypersetup{
    unicode=true,                                           
    plainpages=false,
    pdffitwindow=true,                                      
    colorlinks=true,                                        
    linkcolor=blue,                                        
    citecolor=blue,                                         
    filecolor=blue,                                        
    urlcolor=blue                                          
}

\newcommand*\colvec[3][]{
  
   \begin{pmatrix}\ifx\relax#1\relax\else#1\\\fi#2\\#3\end{pmatrix}
}

\DeclareMathOperator{\Tr}{Tr}

\begin{document}
\title{Photon correlation time-asymmetry and dynamical coherence in multichromophoric systems}
\author{Charlie Nation}
\affiliation{Department of Physics and Astronomy, University College London, Gower Street, WC1E 6BT, London, United Kingdom}
\email{c.nation@ucl.ac.uk}
\author{Hallmann Óskar Gestsson}%
\affiliation{Department of Physics and Astronomy, University College London, Gower Street, WC1E 6BT, London, United Kingdom}
\author{Alexandra Olaya-Castro}
\email{a.olaya@ucl.ac.uk}
\affiliation{Department of Physics and Astronomy, University College London, Gower Street, WC1E 6BT, London, United Kingdom}

\date {\today}

\begin{abstract}
We theoretically investigate polarization-filtered two-photon correlations for the light emitted by a multichromophoric system undergoing excitation transport under realistic exciton-phonon interactions, and subject to continuous incoherent illumination. We show that for a biomolecular aggregate, such as the Fenna-Matthews Olson (FMO) photosynthetic complex, time-asymmetries in the cross-correlations of photons corresponding to different polarizations can be exploited to probe both quantum coherent transport mechanisms and steady-state coherence properties, which are not witnessed by zero-delay correlations. A classical bound on correlation asymmetry is obtained, which FMO is shown to violate using exact numerical calculations. Our analysis indicates that the dominant contributions to time-asymmetry in such photon cross-correlations are population to coherence transfer for Frenkel-Exciton models. Our results therefore put forward photon correlation asymmetry as a promising approach to investigate coherent contributions to excited-stated dynamics in molecular aggregates and other many-site quantum emitters.

\end{abstract}

\maketitle

\textit{Introduction:--}
The direct observation of quantum coherent phenomena in complex systems is a fundamental interdisciplinary scientific challenge that aims to advance our understanding of quantum processes in solid-state \cite{stotz2005, Tran2016, Reigue2017}, biomolecular, and chemical systems \cite{Scholes2011, huelga2013, scholes2017, Polisseni2016, Grandi2019, Mattioni2024}. In these scenarios, the rich interaction of the system of interest with phonon environments makes the task of characterising quantum behaviour challenging. New approaches to probing the influence of coherent mechanisms in the dynamics of such systems, under relevant phonon environments, then promise to open new pathways for the development of robust quantum technologies \cite{Toninelli2021}.

Crucial in each of the aforementioned fields is the development of experimental approaches which may probe effects of quantum coherence in energy transport \cite{hanschke2018, bluvstein2022, Fassioli2010, Scholes2011, Fassioli2012, Chin2013, Oreilly2014, romero2014, romero2017, scholes2017, kim2021, nation2023}. In particular, verifying quantum coherence in photosynthetic excitation transport has been a topic of considerable debate \cite{brixner2005, Olaya-Castro2008,  collini2010a, wilkins2015, holdaway2016, Holdaway2017, duan2017, scholes2017, knee2018, kim2021, werren2023}, which was first given experimental credence by optical multidimensional spectroscopic observations of long-lived `coherence beats' in the two-dimensional electronic spectra of the Fenna-Matthews-Olson (FMO) complex, first at cryogenic temperatures \cite{engel2007}, and later at room temperature \cite{panitchayangkoon2010}. Such nonlinear spectroscopic methods have indeed become the cornerstone of experimental procedures for the investigation of ultrafast excitation dynamics in organic molecules and materials \cite{2Dprimer}.  They however rely on coherent pulsed laser light that, while giving access to a wealth of information, probe molecular systems in regimes outside of their natural conditions, such as weak incoherent irradiation \cite{Fassioli2012}. Furthermore, such methods rely on the optical response of an ensemble of molecules whose properties vary from one to another, making it difficult to discern the quantum processes that are relevant at the single-molecule level.

Since the seminal work of Glauber \cite{Glauber}, photon correlations have become the defining measurements to characterise the quantum nature of light, and are equally pivotal to assess the quantum nature of emitters \cite{Carmichael2,aspect1980, schrama1992, campisi2011, Holdaway2018, DelValle2012, Anderson2018, Velez2019, SanchezMunoz2020, miller2021, landi2023, cygorek2023, downing2023}. More recently photon-counting experiments have been performed on larger and more complex systems \cite{HOLLARS2003393, Hedley2021}, probing even single large supramolecular complexes \cite{Wientjes2014, kim2019, li2023}. The study of photon correlations thus promises new insights into quantum features of single multichromophoric systems, as experimental progress has made near-term study of coherence in energy transport through such methods viable. What is still uncertain are the fundamental signatures of quantum behaviour in multi-chromophoric systems which may be extracted from such experiments.

Recent studies \cite{iles-smith2017, Holdaway2018, cygorek2023, li2023, humphries2023, nation2024} have indeed illustrated the use of second-order photon correlations for the investigation of quantum coherent phenomena. Photon correlations can be bounded under classical assumptions \cite{loudon1980, auffeves2011, SanchezMunoz2020}, and thus a violation of such bounds demonstrates clearly the presence of coherence. 
However, theoretical investigations of photon-correlations that account for the complexity of the electron-phonon interactions characteristic of the systems of interest have just started to emerge \cite{humphries2023, nation2024}. We have previously shown \cite{nation2024} that frequency-filtered and time-resolved photon cross-correlations, and particularly their asymmetry on interchange of measurement order, can be exploited to probe vibronic couplings in a composite emitter system, and the associated coherent dynamical process.

In this work we investigate polarization-filtered photon correlations of the light emitted by a multichromophoric system of several interacting chromophore emitters, each locally coupled to realistic phonon environments. To do this, we deploy the hierarchical equations of motion (HEOM) formalism to open quantum system dynamics and include non-additive incoherent pump and decay channels. We systematically investigate the cases of two- and three-site emitters, and the FMO complex, and show that for small size systems, the violation of classical bounds for the zero-delay photon correlations is remarkably robust to environmental influence. However, since the classical bound grows with the number of coupled emitters, zero-delay photon correlations are not sensitive enough to witness coherence in large systems such as FMO. We then present our main result: time-asymmetry of photon cross-correlations for a multichromophoric system interacting with phonon environments is a sensitive measure to witnesses both coherence properties of the non-steady state as well as the underlying dynamical coherent processes, for which we provide analytical arguments in the supplemental material (SM) \cite{sm}. Notably, there is no single known quantum measure able to capture both steady-state and dynamically induced coherences in quantum processes \cite{Streltsov2017}. The polarization filtered correlations here investigated capture several of the relevant properties of such a measure when applied to quantum transport in multichromophoric systems.

\textit{Set-up:--}
\begin{figure*}
    \includegraphics[width=\textwidth]{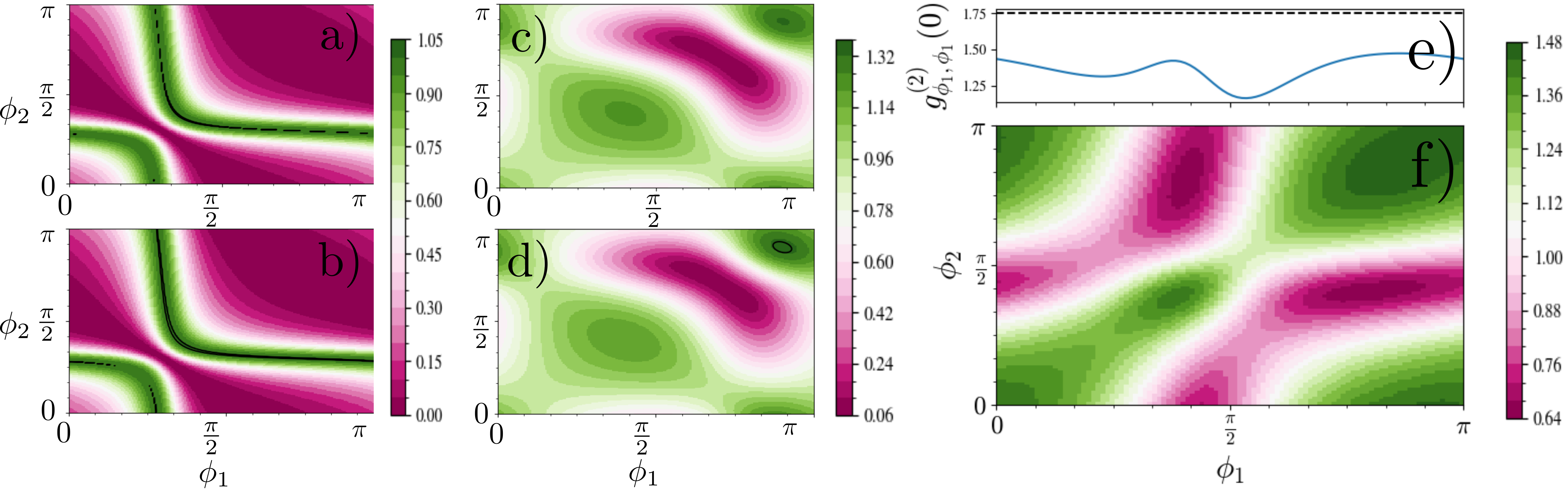}
    \caption{Zero delay photon correlations for different relative filter angles $\phi_1$ and $\phi_2$. a)-b) Show the dimer model with intersite coupling of $J = -46.8$ cm$^{-1}$ (see SM \cite{sm} for additional details) with a) intermediate  environmental coupling $\lambda = 35$ cm$^{-1}$ and b) strong environmental coupling $\lambda = 100$ cm$^{-1}$ (PC645 value). c), d) Show LH2 subunit with c) $\max(J) = 140$ cm$^{-1}$ = $\lambda_1 = \lambda_2$ and d) $\max(J) = 339$ cm$^{-1}$ (LH2 subunit value). f) Zero delay correlations for the FMO complex. e) Shows the autocorrelations for equal polarization angles of FMO (diagonal elements of f)). Additional model details are given in SM \cite{sm}.}
    \label{fig:g20}
\end{figure*}
Collective electronic excitations of molecular aggregates are well described by a Frenkel exciton Hamiltonian \cite{mukamel, Cho2009}, with $M$ chromophores (sites) and a maximal number of $N_{ex} = 2$ excitations,
\begin{align}\label{eq:H_S}
& H_S = \sum_m^M \epsilon_m |m\rangle \langle m | + \sum_{\substack{m,n\\m>n}}^M J_{mn}(|m\rangle \langle n| + |n \rangle \langle m |) \\&
+ \sum_{\substack{m,n\\m>n}}^M(\epsilon_m + \epsilon_n)|m, n\rangle \langle m, n | + \sum_{\substack{m,n, l \\ (m > n) \neq l}}^M J_{mn} ( |n, l\rangle \langle m, l| +  |m, l\rangle \langle n, l| ), \nonumber
\end{align}
with the state $|m\rangle$ denoting a single excitation on site $m$, and $|m, n\rangle$ denoting excitations on sites $m$ and $n$. 
The environment is modelled by a set of local harmonic oscillators with creation (annihilation) operator $b^\dagger_{\xi, m}\, (b_{\xi, m})$ for mode $\xi$ on site $m$, 
$
H_B = \sum_{m, \xi} \omega_{\xi, m} b^\dagger_{\xi, m} b_{\xi, m},
$ each linearly coupled to respective sites via
$
H_{SB} = \sum_m |m\rangle \langle m | \sum_\xi g_{\xi, m} (b_{\xi, m} + b^\dagger_{\xi, m}) + \sum_n \sum_{m \neq n} |m, n\rangle \langle m, n|  \sum_\xi g_{\xi, m} (b_{\xi, m} + b^\dagger_{\xi, m}).
$
We have included the contribution of doubly excited electronic states, without which polarization filtered two-photon correlations trivially vanish at zero-delay. Details of the definition of the second excited manifold for arbitrary operators are given in the SM \cite{sm}.

We capture the influence of the environment via a Drude-Lorentz spectral density \cite{mukamel} for each site,
$
J_m(\omega) = \Theta(\omega) 2\lambda_m \Omega_m \frac{\omega}{\omega^2 + \Omega_m^2},
$
where $\Theta(\omega)$ is the Heaviside step function, $\Omega_m$ is the cutoff frequency, and the reorganisation energy is
$
\lambda_m := \frac{1}{\pi} \int_{-\infty}^{\infty} d\omega J_{m}(\omega) / \omega
$. On-site energies are given by $\epsilon_m = \epsilon_m^0 + \lambda_m$ with $\epsilon_m^0$ the bare electronic excitation energy for site $m$. Exact open system dynamics with the Drude-Lorentz environment is obtained via the HEOM method \cite{ishizaki2006, jin2008, ishizaki2009, tanimura2014a, ke2017}.

Interaction with the electromagnetic environment $H_{EM} = \sum_m \sum_p \omega_{m, p} a^\dagger_{m, p} a_{m, p}$ can support photonic excitations of three polarizations $p \in \{x, y, z\}$, and each couple in the dipole approximation via $H_I = \sum_i \sum_p \sum_m^N g_{i, m}(\hat{\mu}_{m, p} a^\dagger_{i, p} + \hat{\mu}_{m, p}^\dagger a_{i, \lambda})$. The the transition-dipole annihilation operators for polarization $p$ are given by $\hat{\mu}_{p} = \sum_m  |0\rangle \langle m| \sigma_m + \sum_{m > n} (\mu_{a, n} |m\rangle \langle mn| + \mu_{a, m}|n\rangle \langle mn|))$ ,
and may be calculated from the transition-dipole moments $\mu_{m, x},\, \mu_{m, y},\, \mu_{m, z}$ on each site. 
We assume incoherent pumping of the highest energy excitonic state (eigenstate of $H_S$ on the singly excited subspace) at rate $P_{X_1}$, and an incoherent decay of each exciton at rate $\gamma$, unless otherwise stated these are taken to be $\gamma = 1$ ns$^{-1}$ and $P_{X_1} = 0.1$ ns$^{-1}$. We show in the SM \cite{sm} (Eq. \eqref{eq:GKSL-HEOM_fin}) a consistent approach to combining in a non-additive manner these incoherent environmental contributions with the HEOM approach used for the local vibrational environments, in a similar framework to Ref. \cite{fay2022}.

We will analyse three models of the form given above. These are a heterodimer ($M = 2$) model, for which we choose parameters to resemble the central dimer in the photosynthetic antenna phycocyanin 645 \cite{Mirkovic2007a, blau2018a}, with $\lambda = \Omega = 100$ cm$^{-1}$ unless otherwise specified, as well as a subunit of ($M = 3$) of the Light-harvesting 2 (LH2) complex \cite{Cogdell2006, Cupellini2020} found in  purple bacteria. The latter has site-dependent reorganisation energies with the higher energy two sites having $\lambda_1 = \lambda_2 = 140$ cm$^{-1}$, and the lower energy site having $\lambda_3 = 40$ cm $^{-1}$. These simple models will be useful to characterise bounds on zero-delay correlations, where we will see that system size $M$ is of key importance. Our main results are then shown for the full Fenna-Matthews-Olson (FMO) complex \cite{kell2016, hein2012a, kramer2018a, sarovar2010}, with $M = 8$, reorganization energy $\lambda = 35$ cm$^{-1}$ and cutoff $\Omega = 106.1$ cm$^{-1}$. Further model details are shown in the SM. All calculations are performed at 300 K.

Photon correlation functions give information on the conditional probability of time-delayed photon coincidence events \cite{scully_zubairy_1997}. That is, given the detection of a photon with polarization $a$ at $t = 0$, the studied correlations describe the conditional probability of detection of a photon of polarization $b$ at time $t = \tau$. The second-order photon correlation function reads
\begin{align}\label{eq:g2_def}
    g^{(2)}_{ab}(\tau) = \frac{\langle \hat{\mu}_a^\dagger \hat{\mu}_b^\dagger(\tau) \hat{\mu}_b(\tau) \hat{\mu}_a \rangle}{ \langle \hat{\mu}_a^\dagger \hat{\mu}_a\rangle\langle \hat{\mu}_b^\dagger \hat{\mu}_b\rangle},
\end{align}
where $\langle \cdots \rangle := \Tr [\cdots \rho]$, with $\rho$ the steady-state of the system. Assuming the quantum regression theorem, we replace $\hat{\mu}_b^\dagger(\tau) \hat{\mu}_b(\tau) \to (\hat{\mu}_b^\dagger \hat{\mu}_b)(\tau)$, where $O(\tau) = e^{\mathcal{L}^\dagger \tau} O$ is the Heisenberg time evolution of the operator $O$ under the dual map $\mathcal{L}^\dagger$, which is calculated via HEOM as described in the SM. Negative time photon correlations are defined as the reversal of the order of measurements on the system, that is, swapping of the indices $a \leftrightarrow b$. The dynamical map $\mathcal{L}$ is the generator of evolution of the system state, with steady-state $\mathcal{L}[\rho] = 0$. We note the inclusion of the double excited subspace greatly increases the complexity of the HEOM calculations, as the effective state dimension including the ground state is $d = \dim(H) = M(M - 1)/2 + M + 1$. We thus exploit multiple speedups, such as auxiliary-operator filtering \cite{shi2009}, and an efficient steady-state solver \cite{Ye2016} to overcome prohibitive scaling of numerical complexity. 

In simulating the emission of molecular systems \cite{mukamel, adolphs2006, ishizaki2006, hein2012, gelzinis2015}, in Eq. \eqref{eq:g2_def} the operators $\mu_a$ are the transition-dipole operators. In the SM we show how these may be obtained for an arbitrary polarization filter and transition-dipole moments. We show in the SM that over the nanosecond timescale of decay to the electronic ground state, photon correlations indeed tend to their uncorrelated value $g^{(2)}_{ab}(\tau \to \infty) = 1$. Here we focus on picosecond timescales of interaction with the vibrational environment.

We note that current experiments are able to obtain delay time-resolution of a few picoseconds \cite{Korzh2020}, whilst this may limit a complete characterization of time-resolved correlations, the total time-asymmetry is more readily accessible. This is defined as $
    \mathcal{A}[f(t)] = ||f(t) - f(-t)||,
   $
   where $|| f(t) || = \sqrt{h\sum_k f(t_k)}$, $h = \Delta t / N_{points}$, with $\Delta t$ and $N_{points}$ the range in and number of time values, respectively. We observe that this total asymmetry is related to both steady-state coherences and dynamical evolution of coherences in the system, and investigate this looking at the asymmetry over differing timescales. In order to capture the behaviour of steady-state coherences in the system, in the following we compare the asymmetry to the basis independent coherence \cite{Ma2019, le2020}, defined as,
\begin{align}\label{eq:bib}
    C_{1}(\rho) = S(\rho || \mathbb{1}/d ) = \log_2(d) - S(\rho),
\end{align}
where $S(\rho) = - \Tr[\rho \log_2(\rho)]$ is the von Neumann entropy, and $S(\rho||\sigma) = -\Tr[\rho \log(\sigma)] - S(\rho)$ is the relative entropy; $C_{1}(\rho)$ hence measures a distance from the infinite temperature state, and behaves similarly to the purity of the state.

\textit{Zero-delay coherence witness:--}
Zero delay photon-correlations may be exploited to witness the existence of steady-state quantum coherence via the violation of a bound \cite{SanchezMunoz2020}: $g^{(2)}_{ab}(0) \leq 2 - \frac{2}{M}$, which we label the zero-delay bound. Here, in Figure \ref{fig:g20} we show zero-delay polarization filtered correlations for each model. Notably, the structure of bunched/anti-bunched regions is dominated by the molecule-detector orientation. The regions where the zero-delay bound is violated are designated by the inside of the  black solid lines, indicating the presence of steady-state excitonic coherence. 

Figure \ref{fig:g20}a), b) shows the results for the dimer model for two reorganisation energies, with in Figure \ref{fig:g20}b) $\lambda = 100$ cm$^{-1}$, much larger than the electronic coupling of $-46.8$ cm$^{-1}$. We note an increase of the region of violation of this bound when strong environmental coupling is present, which is consistent with the increased excitonic coherences observed as $\lambda$ is increased, shown in SM. In the SM we show that strong Markovian pure dephasing can remove violation of the bound under different pumping conditions as in Ref. \cite{SanchezMunoz2020}. For $M = 2$, violation of the zero-delay bound is robust to the presence of strong environmental coupling, leading to the potential to witness coherence in, for example, fluorescent proteins \cite{Mariano2023, Alejandro} with zero-delay photon cross-correlations.

We note that measurement of the zero-delay photon correlations is experimentally challenging in the presence of fast environmental dephasing \cite{auffeves2011}, as the required time resolution is lower than the dephasing time, typically from hundreds of femtoseconds to a few picoseconds. More easily observed in experiment is then a dephased quasi-equilibrium value, which is typically lower than $g^{(2)}(0)$ \cite{nation2024}, as observed below.

Figure \ref{fig:g20}c), d) show the results for the LH2 subunit \cite{Cupellini2020}, looking at two cases with altered coupling between the most strongly coupled sites: Figure \ref{fig:g20}c) $\max(J) \approx \lambda$ and Figure \ref{fig:g20}d) $\max(J) \gg \lambda$ (note that the latter is the biologically relevant value). We see a region of zero-delay bound violation when $\max(J_{mn}) > \lambda$, however for reduced couplings the bound is no longer violated, despite the presence of steady-state coherences (see Figure \ref{fig:trimer}).

In Figure \ref{fig:g20}e), f) we show zero delay photon cross-correlations for the FMO complex, and observe that the  bound is not violated for any combination of polarization filters, even in the absence of the vibrational environment (see SM). This is due to the fact that the zero-delay bound increases towards the value 2 for large $M$, and thus for multichromophoric systems such as FMO, additional bunching is required for the bound to be violated.  We also note that with knowledge of the dipole moments one can strengthen the bound, and that in this case the bound is similarly not violated (see SM). Thus, as $M$ is increased the zero-delay bound is less sensitive to the presence of steady-state coherence (see Figure \ref{fig:g2t}h)).

\begin{figure}
    \includegraphics[width=0.5\textwidth]{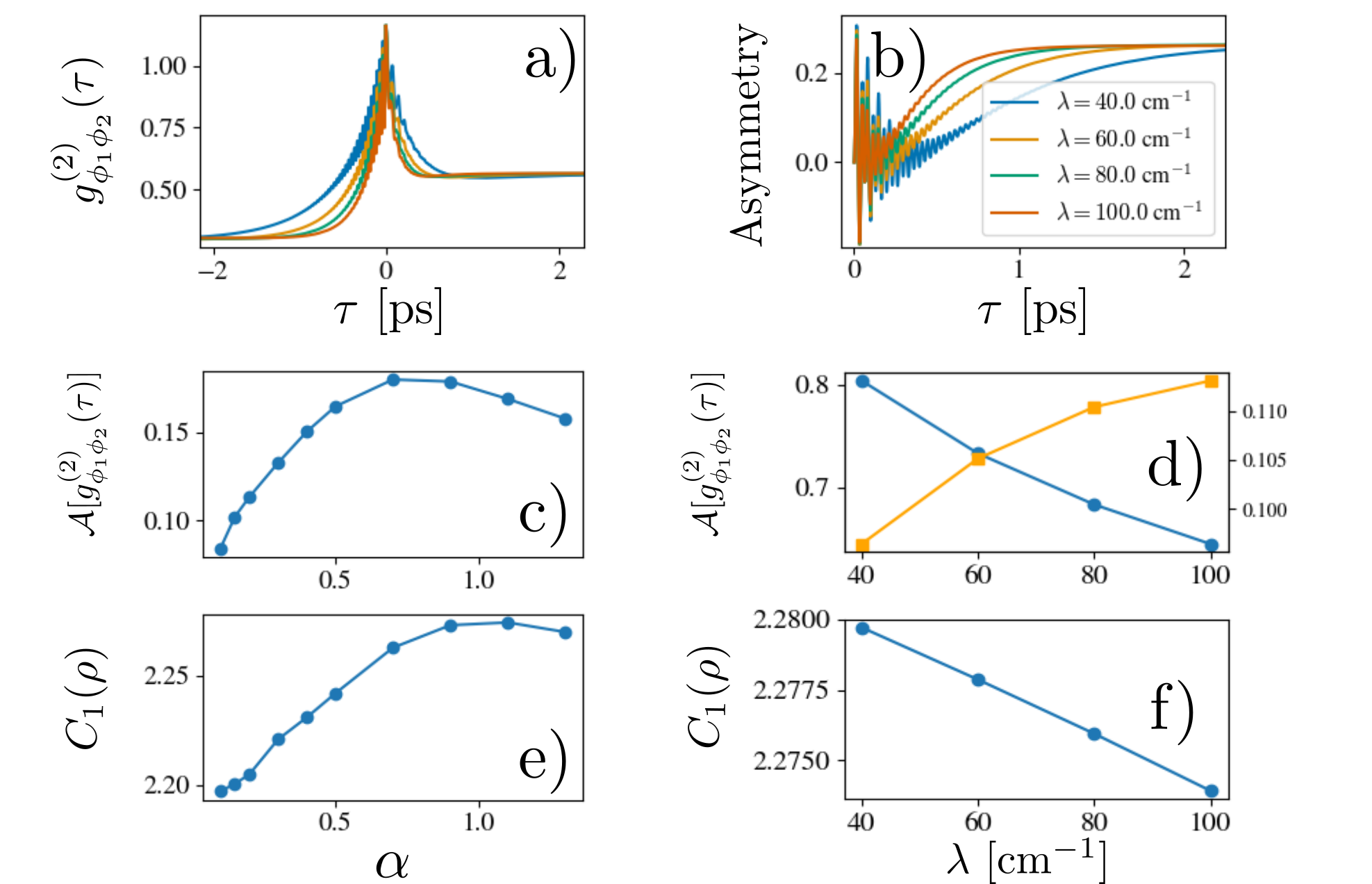}
    \caption{Time-dependence a), b), and total asymmetry c), e) of photon correlations and their comparison to basis independent coherence d), f). e) Shows integration times close to (orange squares, right axis) and larger than (blue circles) the system-environment coherence time $\tau_{S-E} \sim \frac{1}{\sqrt{\Omega \lambda}}$. Dependence on reorganization energy $\lambda$ and coupling strength scaling parameter $\alpha$ shown. HEOM truncation = 20.}
    \label{fig:trimer}
\end{figure}

\begin{figure*}
    \includegraphics[width=\textwidth]{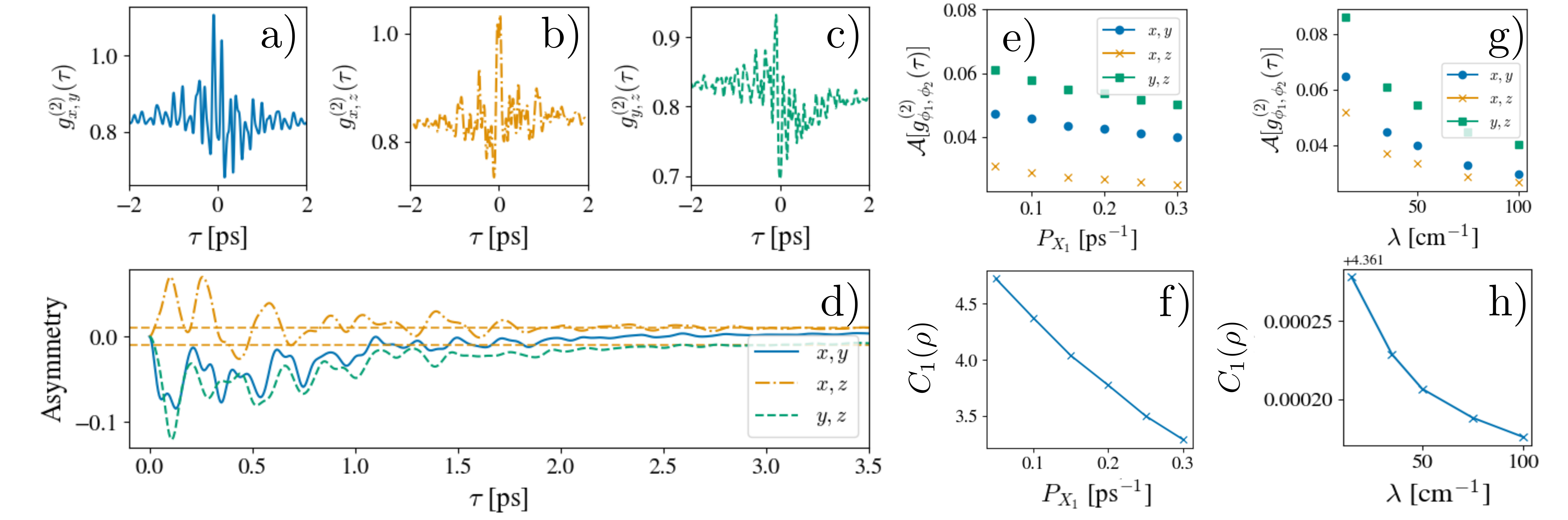}
    \caption{Time dependence of polarization filtered photon correlations for different polarizations a)-c) and their asymmetry for d) for the FMO complex. Dashed yellow line shows the maximal value of the classical bound on correlation asymmetry, which is that of the $x, z$ cross correlation. e), f) Show pump power dependence of the total asymmetry of photon correlations out to 50 ps, and $C_1(\rho)$ (Eq. \eqref{eq:bib}), respectively. g), h) Show the same for dependence on reorganization energy $\lambda$. HEOM truncation is 4, temperature = 300K. We note over nanosecond timescales not shown here each $g^{(2)}_{ab}(\tau) \to 1$.}
    \label{fig:g2t}
\end{figure*}

\textit{Asymmetric photon correlations and quantum coherent transport:--} Our central study is the relation of the behaviour of photon cross-correlations to quantum coherent exciton transfer in photosynthetic biomolecular systems. In the SM (sections II-IV) we provide analytical arguments under the assumption that photon emission processes occur much slower than exciton transfer via an approximation assuming dipole transitions occur predominantly between excitonic states. We show that the dominant sources leading to time-asymmetrical photon cross-correlations are due to excitonic coherences in the steady-state or those generated by early-time Liouvillian dynamics, which we refer to as dynamical coherences. Dynamical coherences are additionally a relevant quantum feature of excitation transport that we wish to probe the influence of in the system. In the following we will not attempt do distinguish contributions to correlation asymmetry from dynamical or steady-state coherence, rather, to probe the presence of either form of coherence. 

We recover symmetric correlations for classical multi-level emitters obeying detailed balance, however, we observe asymmetries may occur in a semi-classical description of coupled two-level systems. These asymmetries can be bounded from the assumptions of an incoherent steady-state, combined with the assumption that excitonic coherences do not contribute to dynamics (see SM, section II), from which we obtain
\begin{align}\label{eq:bound}
    \mathcal{A}[g^{(2)}_{ab}(\tau)] \leq \frac{\sum_{\alpha, \alpha^\prime}| (|\hat{\mu}_{b\alpha^\prime}|^2 |\hat{\mu}_{a\alpha}|^2 - |\hat{\mu}_{a\alpha^\prime}|^2 |\hat{\mu}_{b\alpha}|^2)|}{S^{(1)}_a S^{(1)}_b}p^{(2)},
\end{align}
where $S^{(1)}_a = \Tr[\mu^\dagger_a \mu_a \rho]$. We note this bound is proportional to the population of the doubly excited subspace $p^{(2)}$, and thus expected to be very small. We then analyse the contributions to the Liouvillian of a Frenkel-Exciton model with local pure-dephasing, which may induce asymmetric cross-correlations. We show that in a non-Hermitian Hamiltonian approach the asymmetry inducing processes are those coupling excitonic coherences and populations. Including the influence of quantum jumps, we show that only a single term describing exciton population-population transfer can contribute to photon cross-correlation asymmetry, which appears as a contribution due to excitonic interference enabled by delocalisation. Additional terms which lead to an asymmetry of photon correlations either arise from coherence-coherence or population-coherence coupling terms in the master equation, or coherences in the steady-state, suggesting steady-state and/or dynamical coherence is the dominant source of correlation asymmetry. 

Photon correlation asymmetry can be seen to behave as a witness of coherent transport in Figure \eqref{fig:trimer}, where we study the dependence of the total asymmetry on both electronic couplings and reorganization energy $\lambda$ in the LH2 subunit. We study the dependence on electronic couplings by scaling each coupling strength by a unitless parameter $\alpha$, sending $J_{mn} \to \alpha J_{mn}$ in Eq. \eqref{eq:full_hamil_form}. We see that when altering $\alpha$, both $C_1(\rho)$ and total asymmetry each have very similar dependence. When altering the reorganization energy, however, there more subtle behaviour owing to environmentally induced dynamical coherence at short times. We first note that the excitonic coherences, which increase with $\lambda$, and $C_1(\rho)$, have an opposite dependence on $\lambda$. This is due to the fact that increasing $\lambda$ leads to localisation of the steady-state in the site basis, increasing coherences in the excitonic basis, whereas the basis independent measure captures the decrease in entropy due to this localisation, leading to decreased $C_1(\rho)$. The total asymmetry, on the other hand, has a dependence on $\lambda$ that depends on the timescale of integration. We define a system-environment coherence time as $\tau_{S-E} \sim \frac{1}{\sqrt{\Omega \lambda}}$, over which the system-environment correlations decay \cite{Oreilly2014}. If we choose an integration measure much larger than this coherence timescale (blue circles in Figure \eqref{fig:trimer}e)), we see that the total asymmetry follows a similar behaviour to $C_1(\rho)$. Close to this coherence time (orange squares in Figure \eqref{fig:trimer}e)), we observe that total asymmetry increases with $\lambda$. We thus see that correlation asymmetry at early times is able to capture the increased coherent environmentally induced transport, and thus has concomitant behaviour to pertinent coherence measures in relevant regimes.

In Figure \ref{fig:g2t}a)-d) we show the second-order correlations and their asymmetries for multiple polarization angles for the FMO complex, and observe asymmetries over the expected exciton transport timescales of a few picoseconds in each case \cite{engel2007, kramer2018a} that violate the classical bound of Eq. \eqref{eq:bound}. For simplicity we use the $x, y$ and $z$ component of the dipoles.
In Figure \ref{fig:g2t}e), g) we show how the total asymmetry over a timescale of 50 ps depends on the incoherent pump power and reorganization energy, respectively, noting that pump power is controllable in the laboratory and thus experimentally accessible. We compare these to the basis independent coherence in Figure \ref{fig:g2t}f), h), and observe the same trend in each case as the total asymmetry. The dependence for $C_1(\rho)$ we attribute to the increasing entropy of the state for larger pump power. We thus observe that close to biological conditions photon correlation asymmetry has similar behaviour to $C_1(\rho)$. Indeed, in the SM we show that assuming a maximal entropy steady-state and incoherent dynamics, the total asymmetry is zero, as is $C_1(\rho)$.

\textit{Conclusions:--}
In this work we have shown that photon correlation asymmetry acts as a sensitive probe of quantum coherent energy transport in multichromophoric systems. We have used exact numerical calculations of a prototypical photosynthetic light-harvesting system, the FMO complex, to show that, under realistic electron-phonon interactions, zero delay polarisation-filtered photon correlations may not suffice to witness quantum coherence in non-equilibrium steady-states of such photoexcited system. We have shown in that case correlation time-asymmetries are more reliable reporters of coherence: asymmetry is present over timescales comparable to exciton transfer times at room temperature, and violates a bound obtainable from assuming completely classical dynamics and an incoherent steady state. Further, the total asymmetry is observed to closely follows the behaviour of the basis independent coherence. 

Analytical arguments based on a simplified model show that whilst correlation asymmetry may be induced by population-population transfer terms in a general quantum master equation, for Frenkel-exciton Hamiltonian models with locally coupled environments the dominant sources of asymmetry are excitonic coherences.

Our results thus show that time-resolved photon cross-correlations can be used to probe the quantum coherent nature of transport in single-molecule quantum optical experiments, which can be realised by deploying current on-chip technology \cite{Davanco2017}. This will not only facilitate the much sought after experimental insight into the effect that such coherences may have in the function of photo-active supramolecular complexes, but also the exploration of such complex quantum emitters (and their phonon environments) in the development of next-generation quantum photonic technologies \cite{Toninelli2021}.

\textit{Acknowledgements:--}
We thank the Gordon and Betty Moore Foundation grant GBMF8820 and the Engineering and Physical Sciences Research Council (EPSRC UK) Grant EP/V049011/1 for financial support. We would like to thank Luca Sapienza, Stefano Vezzoli, Thomas Fay, Graham Fleming, Quanwei Li,  K. Birgitta Whaley and Robert L Cook for insightful discussions. The authors acknowledge use of the UCL Myriad High Performance Computing Facility.

\bibliographystyle{apsrev4-1}
\bibliography{bibli}

\newpage

\begin{widetext}
\begin{center}
\begin{titlepage}
  \centering
  \vskip 60pt
  \LARGE  Supplemental Material: \\ Photon correlation time-asymmetry and dynamical coherence in multi-chromophoric systems \par
  \vskip 1.5em
  \large Charlie Nation, Hallmann Óskar Gestsson, and Alexandra Olaya-Castro \par

  \vskip 1.0em
\small \textit{$^{1}$Department of Physics and Astronomy, University College London, Gower Street, WC1E 6BT, London, United Kingdom} \\
  \vskip 0.5em

  \today
    \vskip 2.5em

\end{titlepage}
\end{center}
\setcounter{equation}{0}
\setcounter{figure}{0}
\setcounter{table}{0}
\setcounter{page}{1}
\setcounter{section}{0}

\makeatletter
\renewcommand{\theequation}{S\arabic{equation}}
\renewcommand{\thefigure}{S\arabic{figure}}
\renewcommand{\bibnumfmt}[1]{[S#1]}
\renewcommand{\citenumfont}[1]{S#1}

\section{Second excitation manifold of the Frenkel exciton model}\label{App:second_manifold}
In this section we describe the approach to define the doubly excited manifold, for which we follow Ref. \cite{hein2012} and references therein. We define the matrix elements of the second excited subspace Hamiltonian $H^{(2)}$ in terms of the singly excited subspace $H^{(1)}$ as,
\begin{align}
    \langle mn| H^{(2)} |mn\rangle = \langle m|H^{(1)}|m\rangle +\langle n|H^{(1)}|n\rangle
\end{align}
for diagonal elements, and
\begin{align}
    \langle kl| H^{(2)} & |mn\rangle= 
      \delta_{km}(1 - \delta_{ln})\langle l|H^{(1)}|n\rangle \\ & \nonumber 
    + \delta_{kn}(1 - \delta_{lm})\langle l|H^{(1)}|m\rangle 
    + \delta_{lm}(1 - \delta_{kn})\langle k|H^{(1)}|n\rangle\\ & \nonumber 
    + \delta_{ln}(1 - \delta_{km})\langle k|H^{(1)}|m\rangle.
\end{align}
It will be illustrative below to initially analyse the case where doubly excited states $|mm\rangle$ are included, which amounts to removing the $(1 - \delta_{mn})$ factors in each term above.

The full Hamiltonian including the ground state is then of the form
\begin{align}\label{eq:full_hamil_form}
    H = \begin{pmatrix}
            0 &    0    & 0\\
            0 & H^{(1)} & 0\\
            0 &    0    & H^{(2)}
        \end{pmatrix}.
\end{align}

Then, for an operator $O$ in the second excited subspace we have 
\begin{align}
    O_{ij, ik} = \langle ij |O |ik\rangle = \langle j| O |k \rangle
\end{align}
for $j \neq k$. If we then decompose a dynamical map $M$ into a sum over contributions from individual operators $O$ such that $M = \sum_O M_O$ with $M_O [\cdot]= O_{L} \cdot O_{R}$, where each operator $O_{L/R}$ is defined on the second excited state in the manner above. We then have that
\begin{align}\label{eq:second_space_rates}
    (M_O)_{ij \to ik} &= \langle ik| M_O[|ij\rangle \langle ij|] |ik\rangle \nonumber \\& 
    = \langle ik| O_{L} |ij\rangle \langle ij| O_{R} |ik\rangle \\&
    = \langle k| O_{L} |j\rangle \langle j| O_{R} |k\rangle \nonumber \\& \nonumber 
    = (M_O)_{j\to k}.
\end{align}

\section{Photon correlations for classical maps}\label{sec:proof}
In this section we assert classicality by i) enforcing that the steady-state has no coherences in some basis $|\alpha\rangle$, ii) the generator of dynamics does not create coherences from populations, that is $\mathcal{E}(t)[|\alpha\rangle \langle \alpha |] = \sum_{\beta} W_{\alpha\to \beta}(t) |\beta\rangle \langle \beta|$, where $W_{\alpha\to \beta}$ is understood as the transition probability from state $|\alpha\rangle $ to state $|\beta\rangle$. If this is true of a system, then there exists a basis in which a classical map describes the complete dynamics. Here this basis is taken as the excitonic basis diagonalising $H$, in  which we are interested in probing steady-state coherences, as well as coherences induced by dynamics after photon emission from the steady-state.

In order to calculate the second-order photon correlation functions we must express the dipole operators in terms of the two-excitation subspace of a Frenkel-Exciton model. 
The first simplifying assumption we make in describing photon emission processes is the following:
\begin{align}
    \hat{\mu}_a &= \sum_\alpha(\mu_{a\alpha} |0\rangle \langle \alpha| + \sum_{\beta > \alpha} (\mu_{a\beta} |\alpha\rangle \langle \alpha\beta| + \mu_{a\alpha}|\beta\rangle \langle \alpha\beta|)) \nonumber \\ &
    = \hat{\mu}_a^{(1)} + \hat{\mu}_a^{(2)},
\end{align}
where $\hat{\mu}_a^{(n)}$ acts on the $n$ excitation manifold to map onto that of $n-1$ excitations for $n \in [1, 2]$, and $|\alpha\rangle = \sum_m c_{m\alpha} |m\rangle$ is the excitonic state. The state $|\alpha\beta\rangle$ denotes an excitation on both excitons $\alpha$ and $\beta$. We note that in general, we have $\langle 0| \mu^{(1)}_a |\alpha\rangle = \sum_m \mu_{m,a} \langle m| \alpha \rangle$ and $\langle \alpha| \mu^{(2)} |\beta \gamma\rangle = \sum_{m} \sum_{n\neq m} \mu_m \langle \alpha| m\rangle \langle m n| \beta \gamma\rangle$, and thus the approximation ignores emission processes from the doubly excited subspace of the form $|\alpha, \beta\rangle \to |\gamma\rangle$ for $\gamma \neq \alpha, \, \beta$.

The steady-state in the doubly excited space is
\begin{align}
    \rho = \sum_{\substack{\alpha\beta\alpha^\prime\beta^\prime = 0\\ \alpha > \beta, \alpha\prime> \beta^\prime}} \rho_{\alpha\beta\alpha^\prime\beta^\prime }|\alpha\beta\rangle \langle \alpha^\prime\beta^\prime |.
\end{align}
An incoherent state has the additional assertion that $\alpha= \alpha^\prime$, $\beta = \beta^\prime$. We assume that $\rho$ is the steady-state of the dynamical generator $\mathcal{E}(t)$, such that $\mathcal{E}(t)[\rho] = \rho$. We may understand the second-order photon correlation functions of Eq. \eqref{eq:g2_def} as a sequence of two maps $\Tilde{M}_a[\cdot] = \frac{1}{S^{(1)}_a}  \hat{\mu}_a \cdot \hat{\mu}_a^\dagger = \frac{1}{S^{(1)}_a} M_a[\cdot]$ with $S^{(1)}_a = \langle \hat{\mu}^\dagger_a \hat{\mu}_a \rangle$. Up to a normalisation factor of $\frac{1}{S^{(1)}_aS^{(1)}_b}$ the two-point photon correlation function of Eq. \eqref{eq:g2_def} is obtained via
\begin{align}\label{eq:S2}
    S_{ab}^{(2)}(\tau) := \Tr[M_b[\mathcal{E}(\tau)[M_a[\rho]]]].
\end{align}

We begin by writing an incoherent state in a convenient form:
\begin{align}
    \rho_{inc} = p^{(0)} |0\rangle \langle 0| + \sum_{\alpha=1}^N p^{(1)}_\alpha |\alpha\rangle \langle \alpha| + \sum_{\substack{\alpha\beta \\ \alpha > \beta}}p^{(2)}_{\alpha\beta} |\alpha\beta\rangle \langle \alpha\beta| \nonumber \\  
     = \rho^{(0)} + \rho^{(1)} + \rho^{(2)},
\end{align}
where we have split the contributions into the zero excitation, single excitation, and doubly excited subspaces. We can simply ignore the contribution of the zero excited state to $M_a[\rho]$, as this only has terms $\propto \langle 0 | \alpha\rangle $ for $\alpha \neq 0$, and thus $, M_a[\rho^{(0)}] = 0$. We then have
\begin{align}\label{eq:first_measurement}
M_a[\rho_{inc}] & = \sum_\alpha |\mu_{a\alpha}|^2 p^{(1)}_\alpha |0\rangle \langle 0| \\ &
+ \sum_{\substack{\alpha\beta \\ \alpha > \beta}}p^{(2)}_{\alpha\beta} (|\mu_{a \beta}|^2 |\alpha\rangle \langle \alpha| + |\mu_{a \alpha}|^2 |\beta\rangle \langle \beta| \nonumber \\ & \qquad \qquad 
+ \mu_{a \beta}\mu_{a \alpha}^*  |\alpha\rangle \langle \beta| + \mu_{a \alpha}\mu_{a \beta}^* |\beta\rangle \langle \alpha| ). \nonumber 
\end{align}
We now require the action of $\mathcal{E}$ onto $M_a[\rho_{inc}]$, which can be obtained by defining $\mathcal{E}$ as an incoherent map and thus having the properties:
\begin{align}
    &\mathcal{E}_{inc}(\tau)[|\alpha\beta\rangle \langle \alpha\beta|] = \sum_{\substack{\alpha^\prime\beta^\prime\\\alpha^\prime>\beta^\prime}}W_{\alpha\beta\to \alpha^\prime\beta^\prime}(\tau)|\alpha^\prime\beta^\prime\rangle \langle \alpha^\prime\beta^\prime|  \\ &
    \mathcal{E}_{inc}(\tau)[|\alpha\beta\rangle \langle \alpha^\prime\beta^\prime|] = 0 \quad \textrm{if} \quad \alpha, \beta \neq \alpha^\prime, \beta^\prime.
\end{align}
In this sense we have asserted that both the steady-state and its dynamical generator are `classical', in that they do not contain or generate coherences. We have thus effectively enforced dynamics under a Pauli master equation. 
We then have 
\begin{align}
    \mathcal{E}_{inc} &(\tau)[M_a[\rho_{inc}]] = |\hat{\mu}_{a \alpha}|^2 p^{(1)}_\alpha \mathcal{E}_{inc}(\tau)[|0\rangle \langle 0|] \\ & 
    + \sum_{\substack{\alpha\beta \\ \alpha > \beta}} p^{(2)}_{\alpha\beta} ( |\hat{\mu}_{a \alpha}|^2\mathcal{E}_{inc}(\tau)[|\beta\rangle \langle \beta|] +  |\hat{\mu}_{a \beta}|^2\mathcal{E}_{inc}(\tau)[|\alpha\rangle \langle \alpha|])\nonumber \\ & 
    = \sum_{\alpha=1}^N \sum_{\substack{\alpha^\prime\beta^\prime = 0\\ \alpha^\prime> \beta^\prime}}|\hat{\mu}_{a \alpha}|^2 p^{(1)}_\alpha W_{0 0\to \alpha^\prime\beta^\prime} (\tau)|\alpha^\prime\beta^\prime\rangle \langle \alpha^\prime\beta^\prime| \nonumber \\ & 
    + \sum_{\substack{\alpha\beta = 1 \\ \alpha > \beta}}\sum_{\substack{\alpha^\prime\beta^\prime = 0\\ \alpha^\prime > \beta^\prime}}  p^{(2)}_{\alpha\beta} ( |\hat{\mu}_{a \alpha}|^2 W_{\beta 0 \to \alpha^\prime\beta^\prime}(\tau) \nonumber \\& \qquad \qquad  +  |\hat{\mu}_{a \beta}|^2W_{\alpha0 \to \alpha^\prime\beta^\prime}(\tau))|\alpha^\prime\beta^\prime\rangle \langle \alpha^\prime\beta^\prime . |\nonumber
\end{align}
Here the first term corresponds to an initial de-excitation from the first excited state to the ground state by emission of photon $a$ followed by evolution from $|0 0\rangle \to |\alpha^\prime\beta^\prime \rangle$, including the possibility of no change in the state. The second term, corresponds to photon emission from the doubly excited state to the singly excited state, followed by evolution under $\mathcal{E}_{inc}(\tau)$ to any other state. 

In order to calculate the second-order correlation function, we then wish to find the application of the map $M_b$ onto the above result, for which we find it convenient to re-express the above result into 4 terms corresponding to the transitions that are caused between manifolds by the action of $\mathcal{E}_{inc}(\tau)$. We denote these as terms containing $W^{(01)}_{00\to \alpha^\prime0}$, $W^{(11)}_{\alpha 0\to \alpha^\prime 0}$, $W^{(02)}_{00\to \alpha^\prime\beta^\prime}$, and $W^{(12)}_{\alpha 0\to \alpha^\prime\beta^\prime}$, corresponding to transitions from ground to first excited states, evolution within the first excited state, ground to second excited states, and transitions from the first excited states to the second excited states, respectively. 

We note that typically only the $W^{(11)}_{\alpha 0\to \alpha^\prime 0}$ term will contribute to early time asymmetries, as the transition rate between states of differing excitation number is $\sim 1000$ times smaller than the transition rate between states in the singly excited manifold, and we are interested in the limit of weak incoherent pumping, where the relevant pumping rates are significantly smaller than the decay rates. Thus, we ignore terms that involve transitions between excited state manifolds, as the dominant contribution is that of exciton transfer by $W^{(11)}_{\alpha 0\to \alpha^\prime 0}$.
We additionally exclude any terms that leave the state in the ground state, as these are destroyed by the action of $M_b$. Thus, the relevant terms are

\begin{align}
    \mathcal{E}_{inc}(\tau) &[M_a[\rho_{inc}]]
   \\& \approx  \sum_{\substack{\alpha\beta = 1 \\ \alpha> \beta}}\sum_{\alpha^\prime=1}  p^{(2)}_{\alpha\beta} ( |\hat{\mu}_{a \alpha}|^2 W^{(11)}_{\beta 0 \to \alpha^\prime 0} (\tau) \nonumber \\& \qquad \quad+  |\hat{\mu}_{a \beta}|^2W^{(11)}_{\alpha0 \to \alpha^\prime 0}(\tau))|\alpha^\prime\rangle \langle \alpha^\prime| \nonumber \\ &
   =  \sum_{\substack{\alpha\beta = 1}}\sum_{\alpha^\prime=1}  p^{(2)}_{\alpha\beta} |\hat{\mu}_{a \alpha}|^2 W^{(11)}_{\beta 0 \to \alpha^\prime 0}(\tau)|\alpha^\prime\rangle \langle \alpha^\prime| \nonumber 
\end{align}
where we have extended the definition $p_{\alpha\beta}^{(2)} = p_{\beta\alpha}^{(2)}$ and $p_{\alpha\alpha}^{(2)} = 0$.
 The above form simplifies the calculation of the action of $M_b$, as we are interested only in diagonal terms (due to the trace being taken after $M_b$), which allows us to write

\begin{align}\label{eq:general_classicalG2}
   \Tr[ M_b[  & \mathcal{E}_{inc}(\tau) [M_a[\rho_{inc}]]] ] \nonumber \\ &  \approx \Tr[ M_b^{(1\to0)}[ \mathcal{E}_{inc}^{(1\to1)}(\tau) [M_a^{(2\to1)}[\rho_{inc}]]] ] \\ &  = \Tr[\hat{\mu}_b^{(1)}  \sum_{ij}\sum_{f=1}  p^{(2)}_{\alpha\beta} |\hat{\mu}_{a \alpha}|^2 W^{(11)}_{\beta0 \to \alpha^\prime0}(\tau) |\alpha^\prime\rangle \langle \alpha^\prime| \hat{\mu}_b^{(1)\dagger}]  \nonumber \\&  \nonumber = \sum_{\alpha\beta} \sum_{\alpha^\prime} p_{\alpha\beta}^{(2)} |\hat{\mu}_{b\alpha^\prime}|^2 |\hat{\mu}_{a\alpha}|^2 W^{(11)}_{\beta\to \alpha^\prime} (\tau) \\& \nonumber 
= \sum_{\alpha, \alpha^\prime} |\hat{\mu}_{b\alpha^\prime}|^2 |\hat{\mu}_{a\alpha}|^2 \sum_{\beta} p_{\alpha\beta}^{(2)} W_{\beta\to \alpha^\prime}  (\tau) ,
\end{align}
where the superscript $(n \to n^\prime)$ is used on maps in the first line to represent the terms of the map which take states from the $n$-excitation subspace to the $n^\prime$ excitation subspace. This captures the key approximation implied by the weak pumping and decay condition.

Notably, this expression allows for the total asymmetry to be bounded, as we can write this asymmetry as
\begin{align}
    A &\propto |\sum_{\alpha, \alpha^\prime} (|\hat{\mu}_{b\alpha^\prime}|^2 |\hat{\mu}_{a\alpha}|^2 - |\hat{\mu}_{a\alpha^\prime}|^2 |\hat{\mu}_{b\alpha}|^2) \sum_{\beta} p_{\alpha\beta}^{(2)} W_{\beta\to \alpha^\prime}  (\tau)| \nonumber \\ &
    \leq \sum_{\alpha, \alpha^\prime} |(|\hat{\mu}_{b\alpha^\prime}|^2 |\hat{\mu}_{a\alpha}|^2 - |\hat{\mu}_{a\alpha^\prime}|^2 |\hat{\mu}_{b\alpha}|^2)| \sum_{\beta} p_{\alpha\beta}^{(2)} W_{\beta\to \alpha^\prime}  (\tau),
\end{align}
as $p^{(2)}_{\alpha\beta} > 0$ and $W_{\alpha\to\beta}>0$.
Then using
$\sum_\beta W_{\beta \to \alpha} = \langle \alpha | e^{\mathcal{L^\dagger \tau}} [\sum_\beta|\beta \rangle \langle \beta |]\alpha\rangle = \langle \alpha |e^{\mathcal{L^\dagger \tau}} [\mathbb{1}] |\alpha\rangle  = 1$, where we have used that the adjoint map is unital, which can be seen from the trace preserving condition on the Kraus decomposition of the dynamical map. The Cauchy-Schwarz inequality implies that $\sum_{\beta} p_{\alpha\beta}^{(2)} W_{\beta\to \alpha^\prime}  (\tau) \leq \sum_\beta p_{\alpha\beta}^{(2)} \leq p^{(2)}$, where $p^{(2)}$ is the total occupation of the doubly excited state. This leads us to,
\begin{align}
    A \leq \frac{\sum_{\alpha, \alpha^\prime}| (|\hat{\mu}_{b\alpha^\prime}|^2 |\hat{\mu}_{a\alpha}|^2 - |\hat{\mu}_{a\alpha^\prime}|^2 |\hat{\mu}_{b\alpha}|^2)|}{S^{(1)}_a S^{(1)}_b}p^{(2)}.
\end{align}
 We thus have a classical bound on correlation asymmetry.

\subsection{Examples of correlation symmetry}

We here analyse the case of $N$ multi-level emitters - that is, a classical model as above however where $p_{\alpha\alpha} \neq 0$. In this case, as a consequence of the definition in Section \ref{App:second_manifold} of the doubly excited subspace we have that $\sum_{\beta} p_{\alpha\beta}^{(2)} W_{\beta\to \alpha^\prime} (\tau)  = \sum_\beta p_{\alpha\beta}^{(2)} \langle \alpha^\prime| \mathcal{E}_{inc}(\tau)[|\beta\rangle \langle \beta|]|\alpha^\prime\rangle = \langle \alpha^\prime| \mathcal{E}_{inc}(\tau)[\sum_\beta  p_{\alpha\beta}^{(2)}|\beta\rangle \langle \beta|]|\alpha^\prime\rangle = p_{\alpha\alpha^\prime}$. The final equality here is seen noting that $p^{(2)}_{\alpha\beta} = p_\alpha p_\beta$ for a classical emitter, and thus $\sum_\beta  p_{\alpha\beta}^{(2)}|\beta\rangle \langle \beta| = p_\alpha \sum_\beta p_\beta |\beta\rangle \langle \beta |$, which is proportional to the steady-state of the dynamical map. We then have
\begin{align}\label{eq:final_symmetry}
    \sum_{\alpha, \alpha^\prime} |\hat{\mu}_{b\alpha^\prime}|^2 |\hat{\mu}_{a\alpha}|^2 \sum_{\beta} p_{\alpha\beta}^{(2)} W_{\beta\to \alpha^\prime} (\tau) = \sum_{\alpha, \alpha^\prime} |\hat{\mu}_{b\alpha^\prime}|^2 |\hat{\mu}_{a\alpha}|^2 p_{\alpha \alpha^\prime}^{(2)},
\end{align}
which, crucially, is symmetric on exchange $a\leftrightarrow b$, 
and hence $S_{ab}^{(2)} = S_{ba}^{(2)}$. We thus recover symmetry of correlations in the case of classical multi-level emitters.

In this example we have seen that the relation $p^{(2)}_{\alpha\beta} \propto p_\alpha p_\beta$ leads to the symmetry of photon correlations. This relation is true for example for thermal equilibrium states $\rho_\beta = Z^{-1}\sum_\alpha e^{-\beta E_\alpha}|\alpha\rangle \langle \alpha|$, with $Z = \sum_\alpha e^{-\beta E_\alpha}$.

The above considered case of a state which leads to correlation symmetry leads us to note that for a fully classical system with the assumption of non-invasive measurability, we trivially expect fully time-symmetric correlations of emission. That is, if the initial measurement $M_a$ does not influence the underlying probability distribution: $M_a[\rho] = m_a \rho$ for some value $m_a$.

We can further see from the fact that for weak pumping and decay $\sum_{\beta} W_{\beta \to \alpha} \approx 1$, if the state on the doubly excited manifold is a maximal entropy state $\frac{1}{d^{(2)}} \mathbb{1}$, where $d^{(2)}$ is the dimension on the doubly excited subspace, we recover once more correlation symmetry. This explains why the asymmetry behaves in a similar manner to the basis independent coherence in the limits of high entropy where each are expected to be small.

\subsection{Diagonal Map}

In the above treatment of an incoherent map we completely ignored the presence of coherences, which is a drastic simplification. We can see from Eq. \eqref{eq:first_measurement} that even for an initial state with no coherences, the initial photon emission may lead to a transient state with non-zero coherences. We thus extend the above analysis to the case of a diagonal map, $\mathcal{E}_{D}(\tau)$, where populations map to populations, and coherences to cohernces, such that populations and coherences are completely decoupled. We further saw that the only contribution of the dynamical map in the limit of weak pumping and decay is to the single excited state manifold, thus, we write
\begin{align}
    &\mathcal{E}_{D}(\tau)[|\alpha\rangle \langle \alpha|] = \sum_{\beta}W_{\alpha\to \beta}(\tau)|\beta\rangle \langle \beta|  \\ &
    \mathcal{E}_{D}(\tau)[|\alpha\rangle \langle \beta|] \, = \nonumber \\& \qquad \qquad \sum_{\alpha^\prime \beta^\prime}W^{(C-C)}_{\alpha\beta \to \alpha^\prime \beta^\prime}(\tau)|\alpha^\prime \rangle \langle \beta^\prime| .\nonumber
\end{align}
In the above section we found the term corresponding to the evolution of populations under this map, here we see that the evolution of coherences may also contribute. We obtain, using $M_b^{(1\to 0)}[|\alpha\rangle \langle \beta|] = \mu_{b\alpha} \mu_{b\beta}^* |0\rangle \langle 0|$, that the contribution of this coherence term to photon cross correlations is
\begin{align}
    \sum_{\substack{\alpha\beta \\ \alpha\neq \beta}}   \sum_{\substack{\alpha^\prime\beta^\prime \\ \alpha^\prime\neq \beta^\prime}} p_{\alpha\beta}^{(2)} \mu_{a\alpha} \mu_{a\beta}^*  \mu_{b\alpha^\prime} \mu_{b\beta^\prime}^* W^{(C-C)}_{\alpha\beta \to \alpha^\prime\beta^\prime},
\end{align}
and thus for a diagonal map we have:
\begin{align}
    \Tr[ M_b[  & \mathcal{E}_{D}(\tau) [M_a[\rho_{inc}]]] ]  = \sum_{\alpha, \alpha^\prime} |\hat{\mu}_{b\alpha^\prime}|^2 |\hat{\mu}_{a\alpha}|^2 \sum_{\beta} p_{\alpha\beta}^{(2)} W_{\beta\to \alpha^\prime}  (\tau) \nonumber \\ & + \sum_{\substack{\alpha\beta \\ \alpha\neq \beta}}  \sum_{\substack{\alpha^\prime\beta^\prime \\ \alpha^\prime\neq \beta^\prime}} p_{\alpha\beta}^{(2)} \mu_{a\alpha} \mu_{a\beta}^*  \mu_{b\alpha^\prime} \mu_{b\beta^\prime}^* W^{(C-C)}_{\alpha\beta \to \alpha^\prime\beta^\prime}.
\end{align}
Now, we can see that the coherence-coherence term contributes the asymmetry:
\begin{align}
    &\sum_{\substack{\alpha\beta \\ \alpha\neq \beta}} \sum_{\substack{\alpha^\prime\beta^\prime \\ \alpha^\prime\neq \beta^\prime}} p_{\alpha\beta}^{(2)} \mu_{a\alpha} \mu_{a\beta}^*  \mu_{b\alpha^\prime} \mu_{b\beta^\prime}^* W^{(C-C)}_{\alpha\beta \to \alpha^\prime\beta^\prime} \\ & \nonumber \qquad \qquad - \sum_{\substack{\alpha\beta \\ \alpha\neq \beta}} \sum_{\substack{\alpha^\prime\beta^\prime \\ \alpha^\prime\neq \beta^\prime}} p_{\alpha\beta}^{(2)} \mu_{b\alpha} \mu_{b\beta}^*  \mu_{a\alpha^\prime} \mu_{a\beta^\prime}^* W^{(C-C)}_{\alpha\beta \to \alpha^\prime\beta^\prime} \nonumber \\& =  \sum_{\substack{\alpha\beta \\ \alpha\neq \beta}} \sum_{\substack{\alpha^\prime\beta^\prime \\ \alpha^\prime\neq \beta^\prime}} p_{\alpha\beta}^{(2)}( \mu_{a\alpha} \mu_{a\beta}^*  \mu_{b\alpha^\prime} \mu_{b\beta^\prime}^* - \mu_{b\alpha} \mu_{b\beta}^*  \mu_{a\alpha^\prime} \mu_{a\beta^\prime}^*)W^{(C-C)}_{\alpha\beta \to \alpha^\prime\beta^\prime}.
\end{align}
From which we recover that auto-correlations ($a = b$) are trivially time-symmetric. The above can be re-written by instead relabelling the summed indices:
\begin{align}
 \sum_{\substack{\alpha\beta \\ \alpha\neq \beta}} \sum_{\substack{\alpha^\prime\beta^\prime \\ \alpha^\prime\neq \beta^\prime}} \mu_{a\alpha} \mu_{a\beta}^*  \mu_{b\alpha^\prime} \mu_{b\beta^\prime}^* (p_{\alpha\beta}^{(2)} W^{(C-C)}_{\alpha\beta \to \alpha^\prime\beta^\prime} - p_{\alpha^\prime\beta^\prime}^{(2)} W^{(C-C)}_{\alpha^\prime\beta^\prime \to \alpha\beta}).
\end{align}

\section{Frenkel Exciton with time-dependent pure-dephasing}

Here we show a specific example of an $N$ site Frenkel exciton model with a pure-dephasing environment, as studied extensively in \cite{chin2010, chin2012} in the context of environmentally induced transport processes. Concretely, we have
\begin{align}
    \mathcal{L}[\rho] = -i[H(t), \rho] + \sum_i \frac{\gamma_i(t)}{2}[2|i\rangle \langle i| \rho |i\rangle \langle i| - \{|i\rangle \langle i| , \rho\} ],
\end{align}
where $|i\rangle$ label the sites of the model. In the following, for simplicity, we drop the explicit time dependence on dephasing rates $\gamma_i$ and the Hamiltonian $H(t)$, however note that no assumptions regarding their time dependence are made.

We first wish to write this in the excitonic basis, in order to explore the role of excitonic coherences in the asymmetry of photon correlations. To do so we define $C_{\alpha\beta}^{(i)} := \langle \alpha| i \rangle \langle i | \beta \rangle$, and the master equation becomes
\begin{align}
    \mathcal{L}[\rho] = \bigg( &-i \sum_{\alpha\beta} (E_\alpha \rho_{\alpha\beta} - E_\beta \rho_{\alpha\beta}) \nonumber \\& - \sum_i \frac{\gamma_i}{2} \sum_{\alpha\beta\alpha^\prime} (C_{\alpha\alpha^\prime}^{(i)} \rho_{\alpha^\prime\beta}  + C_{\alpha^\prime\beta}^{(i)} \rho_{\alpha\alpha^\prime}) \nonumber \\ & 
    + \sum_i \gamma_i \sum_{\alpha\beta\alpha^\prime\beta^\prime}C_{\alpha\beta^\prime}^{(i)}C_{\alpha^\prime\beta}^{(i)}\rho_{\beta^\prime \alpha^\prime}  \bigg) |\alpha\rangle \langle \beta|.
\end{align}
The first two terms of the above master equation can be written in terms of a non Hermitian Hamiltonian $H_{NH} = H - i \sum_i \frac{\gamma_i}{2} |i\rangle \langle i| = H - i \sum_i \frac{\gamma_i}{2} \sum_{\alpha\beta} C^{(i)}_{\alpha\beta} |\alpha\rangle \langle \beta |$, by defining $\mathcal{L}_{NH}[\cdot] = -i(H_{NH}\cdot - 
 \cdot H_{NH}^\dagger)$. The third term is that due to quantum jumps, which we write as $\mathcal{L}_J$.

We have seen in the above that the asymmetry of photon correlations can be analysed in four steps. 1) Initial measurement on steady-state $\rho^{(a)} = M_a^{(2\to1)}[\rho]$. 2) Time evolution under $\mathcal{L}$. 3) Second measurement $M_b^{(1\to0)}[\rho^{(a)}(t)]$. 4) Analysis of symmetry under interchange of measurement order via that of indexes $a, b$. In order to determine the symmetry, for simplicity, in step 2) rather than calculating the full time evolution we can calculate the application of the dynamical generator, or Liouvillian $\mathcal{L}$.

We first write the steady-state of the system, which we assume to be incoherent, as 
\begin{align}
    \rho_{inc} & = p^{(0)} |0\rangle \langle 0| + \sum_{\alpha=1}^N p^{(1)}_\alpha |\alpha\rangle \langle \alpha| + \sum_{\substack{\alpha\beta \\ \alpha > \beta}}p^{(2)}_{\alpha\beta} |\alpha\beta\rangle \langle \alpha\beta| \nonumber \\  &
     = \rho^{(0)} + \rho^{(1)} + \rho^{(2)}.
\end{align}

Beginning with step 1), then, we apply the measurement via emission from the doubly excited subspace to the single excited subspace:
\begin{align}\label{eq:init_measurement}
    \rho^{(a)}_{inc} = \sum_{\substack{\alpha\beta \\ \alpha \neq \beta}} p^{(2)}_{\alpha\beta} (|\mu_{a\alpha}|^2 |\beta\rangle \langle \beta| + \mu_{a\alpha}\mu_{a\beta}^*|\alpha\rangle \langle \beta|).
\end{align}
We thus observe that the initial emission process creates coherences from an initial incoherent state.

For step 2) we then calculate the action of the Liouvillian $\mathcal{L}$, which we split as above into two parts. First, the non-Hermitian Hamiltonian part, which acts on matrix elements $|\alpha \rangle \langle \beta|$ to obtain 
\begin{align}
    \mathcal{L}_{NH}[|\alpha\rangle \langle \beta |] & = -i \sum_{\alpha^\prime}  ( |\alpha^\prime\rangle \langle \alpha^\prime |H_{NH} |\alpha\rangle \langle \beta| -  |\alpha\rangle \langle \beta| H_{NH}^\dagger |\alpha^\prime\rangle \langle \alpha^\prime|)\nonumber  \\& 
    = -i \sum_{\alpha^\prime} ((E_\alpha \delta_{\alpha\alpha^\prime} - i\sum_i \frac{\gamma_i}{2} C^{(i)}_{\alpha^\prime\alpha})|\alpha^\prime\rangle \langle \beta| \\& \nonumber \qquad  \qquad - (E_\beta \delta_{\beta\alpha^\prime} + i\sum_i \frac{\gamma_i}{2} C^{(i)}_{\beta \alpha^\prime})|\alpha\rangle \langle \alpha^\prime|) \\& \nonumber 
    = \sum_{\alpha^\prime} ((E_\alpha \delta_{\alpha\alpha^\prime} - i\Gamma_{\alpha^\prime\alpha})|\alpha^\prime\rangle \langle \beta|  \nonumber \\& \qquad \qquad - (E_\beta \delta_{\beta\alpha^\prime} + i\Gamma_{\beta \alpha^\prime})|\alpha\rangle \langle \alpha^\prime|),
\end{align}
where we have defined $\Gamma_{\alpha^\prime\alpha} = \sum_i \frac{\gamma_i}{2} C^{(i)}_{\alpha^\prime\alpha} $. Acting on $\rho^{(a)}_{inc}$, we thus recover
\begin{align}
    \mathcal{L}_{NH}[\rho^{(a)}_{inc}] = & 
  -\sum_{\substack{\alpha\beta \\ \alpha \neq \beta}} p^{(2)}_{\alpha\beta}|\mu_{a\alpha}|^2 \sum_{\alpha^\prime}( \Gamma_{\alpha^\prime \beta}|\alpha^\prime\rangle \langle \beta|  - \Gamma_{\beta \alpha^\prime}|\beta\rangle \langle \alpha^\prime|) \nonumber \\ &
 -i \sum_{\substack{\alpha\beta \\ \alpha \neq \beta}} p^{(2)}_{\alpha\beta} \mu_{a\alpha}\mu_{a\beta}^* \sum_{\alpha^\prime} \bigg[(E_\alpha \delta_{\alpha\alpha^\prime} - i\Gamma_{\alpha^\prime\alpha})|\alpha^\prime\rangle \langle \beta| ) \nonumber \\ &   \qquad \qquad - (E_\beta \delta_{\beta\alpha^\prime} + i\Gamma_{\beta \alpha^\prime})|\alpha\rangle \langle \alpha^\prime|)\bigg].
\end{align}
Similarly, for the jump term, we have
\begin{align}
    \mathcal{L}_J[|\alpha\rangle \langle \beta| ] = & \sum_i \gamma_i \sum_{\alpha^\prime \beta^\prime} C_{\alpha^\prime \alpha}^{(i)} C_{\beta\beta^\prime}^{(i)} |\alpha^\prime \rangle \langle \beta^\prime|,
\end{align}
such that
\begin{align}
    \mathcal{L}_J[\rho^{(a)}_{inc}] =  \sum_{\substack{\alpha\beta \\ \alpha \neq \beta}} p^{(2)}_{\alpha\beta} &  \sum_i \gamma_i \sum_{\alpha^\prime \beta^\prime} ( |\mu_{a\alpha}|^2 C_{\alpha^\prime \beta}^{(i)} C_{\beta\beta^\prime}^{(i)}  \\& + \mu_{a\alpha}\mu_{a\beta}^* C_{\alpha^\prime \alpha}^{(i)} C_{\beta\beta^\prime}^{(i)} ) |\alpha^\prime \rangle \langle \beta^\prime| \nonumber .
\end{align}
Now, for step 3), we calculate the second measurement process, given by the emission to the ground state via $M_b^{(1\to 0)}$. This can similarly be seen to act on matrix elements on the single excited space to yield
\begin{align}
    M_b^{(1\to 0)}[|\alpha\rangle \langle \beta |] = \mu_{b\alpha} \mu_{b\beta}^* |0\rangle \langle 0 |.
\end{align}
We thus have, for the non-Hermitian contribution,
\begin{align}\label{eq:non-Hermitian_part}
    \Tr[M_b^{(1\to 0)}[& \mathcal{L}_{NH}[\rho^{(a)}_{inc}]]] = \nonumber \\&
  -\sum_{\substack{\alpha\beta \\ \alpha \neq \beta}} p^{(2)}_{\alpha\beta}|\mu_{a\alpha}|^2 \sum_{\alpha^\prime}( \mu_{b\alpha^\prime} \mu_{b\beta}^* \Gamma_{\alpha^\prime \beta} - \mu_{b\beta} \mu_{b\alpha^\prime}^*   \Gamma_{\beta \alpha^\prime}) \nonumber \\ &
 -i \sum_{\substack{\alpha\beta \\ \alpha \neq \beta}} p^{(2)}_{\alpha\beta} \mu_{a\alpha}\mu_{a\beta}^* \sum_{\alpha^\prime} (E_\alpha \delta_{\alpha\alpha^\prime} - i\Gamma_{\alpha^\prime\alpha})  \mu_{b\alpha^\prime} \mu_{b\beta}^* ) \nonumber \\ &   \qquad \qquad - (E_\beta \delta_{\beta\alpha^\prime} + i\Gamma_{\beta \alpha^\prime}) \mu_{b\alpha} \mu_{b\alpha^\prime}^*)
\end{align}
and secondly, for the quantum jump term
\begin{align}\label{eq:Jump_part}
    \Tr[ M_b^{(1\to 0)}[\mathcal{L}_{J}[\rho^{(a)}_{inc}]]] =  &  \sum_{\substack{\alpha\beta \\ \alpha \neq \beta}} p^{(2)}_{\alpha\beta} \sum_i \gamma_i \sum_{\alpha^\prime \beta^\prime} ( |\mu_{a\alpha}|^2 C_{\alpha^\prime \beta}^{(i)} C_{\beta\beta^\prime}^{(i)} \nonumber  \\& + \mu_{a\alpha}\mu_{a\beta}^* C_{\alpha^\prime \alpha}^{(i)} C_{\beta\beta^\prime}^{(i)} ) \mu_{b\alpha^\prime} \mu_{b\beta^\prime}^*.
\end{align}

Finally, then, we can observe in each of these contributions which terms lead to symmetric photon correlations, and which lead to an asymmetry on interchange of indices $a, b$. We crucially wish to distinguish the role of terms which couple populations to populations, those which couple coherences to coherences, and those coupling populations and coherences. 

First analysing the non-Hermitian term, Eq. \eqref{eq:non-Hermitian_part}, we can note that the first term manifests from the populations in $\rho^{(a)}_{inc}$, and consists of terms in the summation coupling these populations to both populations and coherences. This can be seen via noting that
\begin{align}\label{eq:non-hermitian_term_1}
    p^{(2)}_{\alpha\beta}|\mu_{a\alpha}|^2 \mu_{b\alpha^\prime} \mu_{b\beta}^* \Gamma_{\alpha^\prime \beta} 
\end{align}
arises from the matrix element $p^{(2)}_{\alpha\beta}|\alpha\beta\rangle \langle \alpha \beta| $ of the initial state  being acted upon by the initial measurement $M_a^{(2\to 1)}$ to obtain $p^{(2)}_{\alpha\beta}|\mu_{a\alpha}|^2|\beta\rangle \langle \beta| $. The non-Hermitian part of the generator then acts on this state, the first term of which includes $\Gamma_{\alpha^\prime \beta} = \sum_i \frac{\gamma_i}{2} C^{(i)}_{\alpha^\prime \beta}$ which arises due to the non-Hermitian part of the effective Hamiltonian acting from the left onto the population $\propto |\beta\rangle \langle \beta|$. This term thus maps the population $|\beta\rangle \langle \beta|$ to itself for $\alpha^\prime = \beta$, and otherwise maps populations to coherences in the excitonic basis. We note that the term that maps the population to itself, $ p^{(2)}_{\alpha\beta}|\mu_{a\alpha}|^2 | \mu_{b\beta}|^2 \Gamma_{\beta \beta} $, is symmetric under interchange of indices $a$ and $b$, and thus is symmetric with respect to reversal of photon measurement order. The terms which map populations to coherences in the dynamics contribute the asymmetry in Eq. \eqref{eq:non-hermitian_term_1}.

The second term (lines 2 and 3 of the right hand side) of Eq. \eqref{eq:non-Hermitian_part} arises from the coherences generated by the initial measurement  $M_a^{(2\to 1)}$, and contains terms of the form
\begin{align}\label{eq:non-hermitian_term_2}
    p^{(2)}_{\alpha\beta} \mu_{a\alpha}\mu_{a\beta}^* \sum_{\alpha^\prime} (E_\alpha \delta_{\alpha\alpha^\prime} - i\Gamma_{\alpha^\prime\alpha})  \mu_{b\alpha^\prime} \mu_{b\beta}^* .
\end{align}
Here we see that the generated coherence $\propto |\alpha\rangle \langle \beta |$ with $\alpha\neq \beta$ under the initial measurement $M_a$ interacts (in this case again from the left) with the non-Hermitian Hamiltonian. The Hamiltonian part maps eigenstates to themselves, and thus contributes $E_\alpha \delta_{\alpha\alpha^\prime}$, and thus this term maps coherences to coherences. Notably the term $ p^{(2)}_{\alpha\beta} E_\alpha \mu_{a\alpha}\mu_{a\beta}^*  \mu_{b\alpha} \mu_{b\beta}^* $ is symmetric on exchange of indices $a$ and $b$. 

The second term in Eq. \eqref{eq:non-hermitian_term_2} includes elements which map coherences to coherences, and coherences to populations. The former include each term in Eq. \eqref{eq:non-hermitian_term_2} with $\alpha^\prime \neq \alpha$, which can be seen from the corresponding term of Eq. \eqref{eq:non-Hermitian_part}. Notably, the term $p^{(2)}_{\alpha\beta }  \Gamma_{\alpha^\prime\alpha} \mu_{a\alpha}\mu_{a\beta}^* \mu_{b\alpha^\prime} \mu_{b\beta}^*$ is symmetric under interchange of indices $a$, $b$ only for $\alpha^\prime = \alpha$ - the term mapping coherences to coherences. Again, those mapping populations to coherences manifest correlation asymmetry.

We thus see that terms in the non-Hermitian effective Hamiltonian that lead coherences and populations to evolve separately do not lead to correlation asymmetry, whereas terms coupling populations and coherences can lead to asymmetric photon correlations. 

We now turn to the quantum jumps term of the master equation. Taking as an example the first term in \eqref{eq:Jump_part}, which describes the evolution and subsequent emission event of the populations term in Eq. \eqref{eq:init_measurement}, the factor $C_{\alpha^\prime \beta}^{(i)} C_{\beta\beta^\prime}^{(i)}$ arises from the jump operator acting on the population $|\beta\rangle \langle \beta|$, which is mapped to $|\alpha^\prime\rangle \langle \beta^\prime|$, and thus this term maps populations to populations when $\alpha^\prime = \beta^\prime$, and otherwise maps populations to coherences. We note that when $\alpha^\prime = \beta^\prime$ the relevant term is $ p^{(2)}_{\alpha\beta} \gamma_i |\mu_{a\alpha}|^2 C_{\alpha^\prime \beta}^{(i)} C_{\beta\alpha^\prime}^{(i)} |\mu_{b\alpha^\prime}|^2$, which is not in general symmetric under interchange of $a$ and $b$, meaning that the quantum jumps term mapping populations to populations may indeed generate asymmetries in photon correlations. Here symmetric correlations are recovered in the case of no exciton delocalisation, when $C_{\alpha^\prime\beta}^{(i)} = \delta_{\alpha^\prime \beta}$. 
We further note that the term $C_{\alpha\beta}^{(i)}C_{\beta\alpha}^{(i)} = \langle \beta |i\rangle \langle i| \alpha \rangle\langle \alpha |i\rangle \langle i| \beta \rangle$ can be understood in terms of interferences between excitons $\alpha$ and $\beta$ on site $i$. For $\alpha = \beta$ the sum over each site corresponds to the inverse participation ratio of the excitonic state, capturing the delocalisation of the exciton over all sites. 

The second term in Eq. \eqref{eq:Jump_part} describes the evolution by quantum jumps of the coherences formed via the initial measurement of $M_a^{(2\to 1)}$. This term maps coherences to populations for $\alpha^\prime = \beta^\prime$, and coherences to coherences otherwise. The relevant factor for this term is given by $p^{(2)}_{\alpha\beta} \gamma_i \mu_{a\alpha}\mu_{a\beta}^* C_{\alpha^\prime \alpha}^{(i)} C_{\beta\beta^\prime}^{(i)} \mu_{b\alpha^\prime} \mu_{b\beta^\prime}^*$, which is symmetric for real $\mu$ in the  $\alpha^\prime = \beta^\prime$ case, and otherwise only symmetric when this condition is again enforced in the limit of zero exciton delocalisation. We note that for Hamiltonians describing time-reversal invariant systems the eigenstates are real valued, and thus $\mu$ is similarly real in the exciton basis. This occurs for the Frenkel exciton Hamiltonian in the cases studied where all couplings are real valued.

We thus see that the terms which may lead to a violation of the symmetry on interchange of measurement order either manifest as a result of the mapping of coherences to populations and vice-versa, or arise in the quantum jumps term as a result of exciton delocalisation. This thus leads to the conclusion that correlation asymmetry in Frenkel exciton models with environmental couplings locally acting in the site basis is a direct result of manifestly quantum behaviours of the system.

\section{Models}
In this section we describe in full the models used in the main text.
Each model consists of three components required to calculate the steady-state and time evolution of photon cross correlations: i) the Hamiltonian, ii) the environment as characterised by a spectral density and/or jump operators with respective rates, iii) the dipole moments, which may be used to obtain their projection onto a polarization direction as shown in Section \ref{app:project_mu}, and obtain the dipole operators $\hat{\mu}$. All calculations are performed at room temperature, 300 K.

\subsection{Dimer}
We choose the dimer model parameters to resemble PC645 \cite{blau2018a}. We thus choose site energies as $\epsilon^0_1 = 17317$ cm$^{-1}$, $\epsilon^0_1 = 15405 $ cm$^{-1}$, and the coupling strength as $J = -46.8 $  cm$^{-1}$ (when not specified), and with a Drude-Lorentz environment with $\lambda = \Omega = 100$  cm$^{-1}$.

We take the dipole moment of the dimer model to be given by a single angle $\theta$ dictating the relative orientation of the moments. Facing on in the molecular $x$ direction, and the other an angle $\theta$ in the $x-y$ plane we then have $\vec{\mu}_1 = (\mu, \, 0,\, 0)$ and $\vec{\mu}_2 = (\mu \cos(\theta),\, \mu \sin(\theta), \, 0)$. We take $\mu = 1$ for simplicity, and use $\theta = \pi / 2$ in the calculations shown. The relative angle of the molecular and lab frames is further specified by the angles $\theta_{det},\, \phi_{det}$ (see Sec \ref{app:project_mu} below), which we take to be equal to 0.8 $\pi$ and $0.4 \pi$, respectively. We note that it is these values as well as the relative dipole orientations which dictate the overall shape of the cross-correlation colour plots in e.g Figure \ref{fig:g20}. Thus we do not observe the shape change significantly with Hamiltonian parameters, but rather the magnitude of correlations.

We note that, as a naively applied dimer model we do not expect the direct results of this model to in practice resemble PC645, which itself has 8 sites and multiple environmental modes which play an important role in exciton dynamics, which are not considered here. This model is chosen as a simple model with similar parameters to those expected in photosynthetic systems.

In Figure \ref{fig:g20SMS} we use different parameters chosen to be identical to those in Ref. \cite{SanchezMunoz2020}, and add an additional pure dephasing term to the environment via a Gorini, Kossakowski, Lindblad and Sudarshan (GKSL) master equation with jump operators $L_{pd} = |m\rangle \langle m|$ acting on each site with an identical rate $\gamma_{pd}$.

\subsection{LH2 subunit}

The subunit for which we show results for zero-delay photon correlations in Figures \ref{fig:g20}c), d) in the main text is chosen as a single $\alpha-\beta$ subunit of LH2, consisting of a pair of chromophores from the B850 ring, which consists of 18 total sites, and a single chromophore of the B800 ring, which is made up of 9 sites. This subunit is thus repeated 9 times in a ring structure in a single LH2 complex. The parameters we use for this model are taken from Ref. \cite{Cupellini2020}, which we repeat here for clarity. 

The Hamiltonian is given by
\begin{align}
H = 
    \begin{pmatrix}
12798.4 & 339.0& -15.6 \\
339.0 & 12805.8 & -9.7  \\
-15.6 & -9.7 & 13021
\end{pmatrix} ,
\end{align}
where the upper two diagonal elements are the (strongly coupled) $\alpha$ and $\beta$ sites of the B850 ring, and the third is the B800 site, which is far more weakly coupled. The Drude-Lorentz environments of the B850 chromophores, sites 1 and 2, each have reorganization energy $\lambda_1 = \lambda_2 = 140$ cm$^{-1}$ and cutoffs $\Omega_1 = \Omega_2 = 100$ cm$^{-1}$ , whereas the B800 chromophore has a lower reorganization energy $\lambda_3 = 40$ cm$^{-1}$, and an identical cutoff $\Omega_3 = 100$ cm$^{-1}$. In Figures \ref{fig:g20}c) we reduce the coupling $J_{12}$ from 339.0 cm$^{-1} \gg \lambda_1$ to 140 cm$^{-1} = \lambda_1$. The dipole moments are $\vec{\mu}_{1} = (8.7509, 2.9632, 1.1907)$, $\vec{\mu}_{2} = (4.3143, - 7.4567, 0.842)$, and $\vec{\mu}_{3} = (-5.6751, 6.3565, 1.2896)$, respectively. In Figures \ref{fig:g20}c), d) we use $\theta_{det} = \pi / 2$ and $\eta_{det} = \pi / 20$, which are chosen to show the area in which the zero-delay bound is violated in Figures \ref{fig:g20}d).

\subsection{FMO}

The Hamiltonian used is `Model C' of Ref. \cite{kell2016}:
\begin{widetext}
\begin{equation*}\label{eq:PC645_Hamil}
H_{FMO} = 
\begin{pmatrix}
12405.& -87& 4.2& -5.2& 5.5& -14.& -6.4& 21. \\
 -87& 12505.& 28.& 6.9& 1.5& 8.7& 4.5& 4.2\\
 4.2&  28.& 12150.& -54.& -0.2& -7.6& 1.2& 0.6\\
 -5.2&  6.9&  -54.& 12300.& -62.& -16.& -51.& -1.3\\
 5.5& 1.5  & -0.2& -62. & 12470.& 60.& 1.7& 3.3\\
  -14.& 8.7 &  -7.6& -16. & 60. & 12575.& 29.& -7.9\\
 -6.4& 4.5 & 1.2 &  -51.&  1.7& 29. & 12375.& -9.3\\
 21. & 4.2 &  0.6 & -1.3 &  3.3& -7.9 & -9.3 & 12430
\end{pmatrix}.
\end{equation*}

\begin{table}
    \centering
        \caption{Components of dipole moments of FMO complex \cite{kramer2018a} can be calculated from the protein data bank (code: 3ENI).}
    \begin{tabular}{|c|c|c|c|c|c|c|c|c|} \hline 
            & 1&  2&  3&  4&  5&  6&  7& 8\\ \hline 
         x&-0.74100554&  -0.85714086&  -0.19712114 &  -0.79924043&  -0.73692547&  -0.13501747&  -0.49511476& -0.13838472\\ \hline 
       y&  -0.56060174&   0.5037757&   0.95741018&  -0.53357321&  0.6557619 &   -0.879218&  -0.70834118& 0.82141214 \\ \hline 
     z&    -0.36964371&  -0.10732938&  -0.21097155&   -0.27661231&   0.16406458&  0.45688729&  -0.50310451& -0.55329175\\ \hline
    \end{tabular}
    \label{tab:FMO_dips}
\end{table}
\end{widetext}

As in Ref. \cite{kell2016} the employed form of the environment is not amenable to description in terms of the HEOM formalism, we instead use a Drude-Lorentz form with $\lambda = 35$cm$^{-1}$ and $\Omega = 106.1$ cm$^{-1}$ as in Refs. \cite{hein2012a, kramer2018a} which is also similar to that used/observed in \cite{sarovar2010}. The dipole moments are assumed to lie in the direction of a line connecting the NB and ND atoms. The atomic coordinates were taken from the crystal structure of the FMO trimer. Dipole moments are shown in Table \ref{tab:FMO_dips}.

In our calculations for the time dependent correlations FMO complex we have used the molecular $x, \, y,\, $ and $z$ components for simplicity, noting that these are possible to select from combinations of polarization angle and molecular orientation, though more importantly that qualitative features with different selections of these parameters are not altered. For the zero delay photon statistics, we use the same relative molecule-detector orientations as the dimer model of $\theta_{det} = 0.8 \pi$ and $\phi_{det} = 0.4\pi$.

\section{Projection of molecular dipole moment onto effective polarization-filter}\label{app:project_mu}

\begin{figure}
    \includegraphics[width=0.45\textwidth]{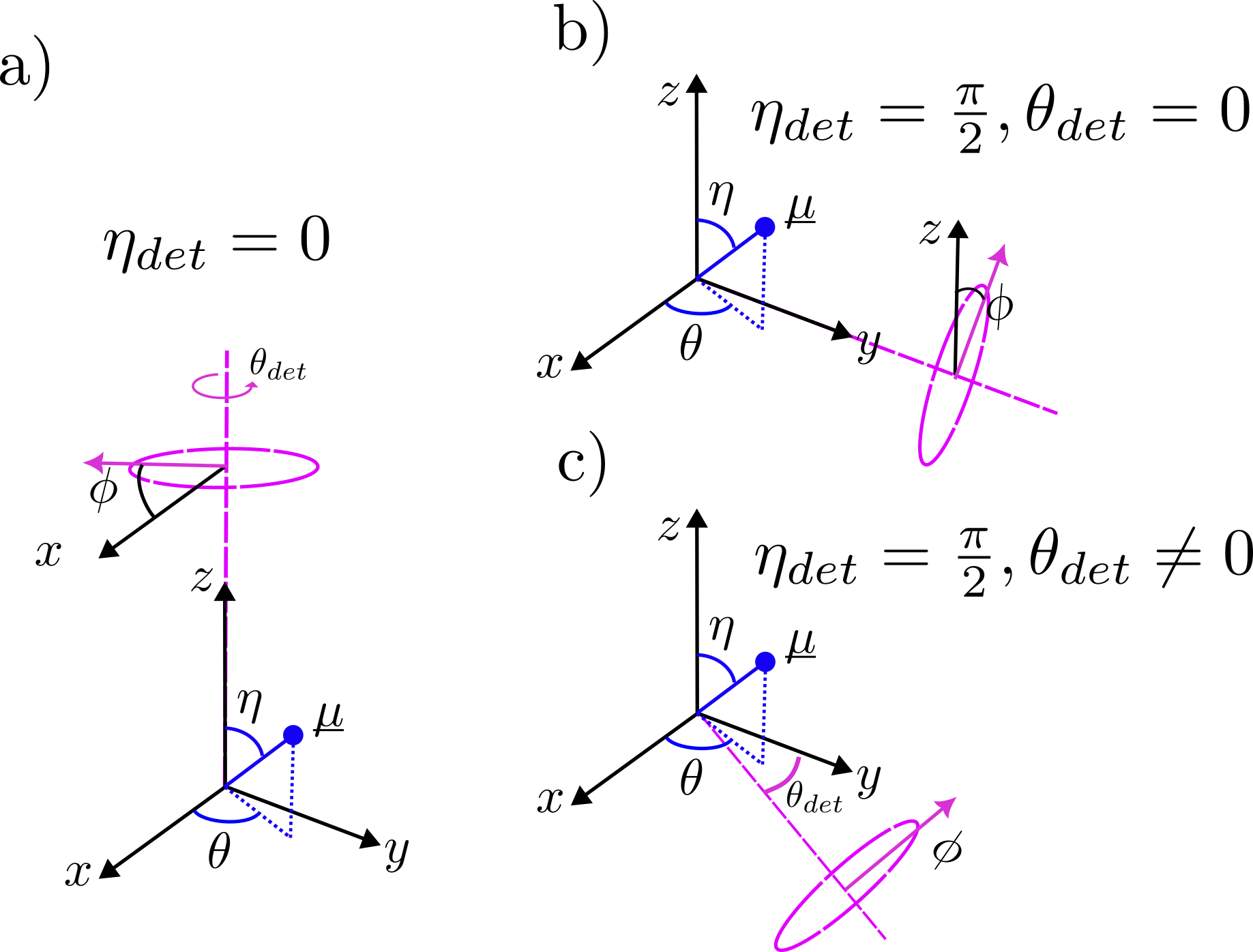}
    \caption{Examples of relative molecule-detector orientations for obtaining effective polarization-filtered dipole moment.}
    \label{fig:effective_dipoles}
\end{figure}

The dipole moment of site $i$ is defined as a three component vector $\vec{\mu}_m = (\mu_{m, x}, \, \mu_{m, y}, \, \mu_{m, z})$. This vector can be defined in terms of its magnitude $\mu_m = |\vec{\mu}_m| = \sqrt{\mu_{m, x}^2 +\mu_{m, y}^2 + \mu_{m, z}^2}$, and the azimuthal and polar angles, $\eta = \arccos(\frac{\mu_{z, i}}{\mu})$ and $\theta = \arctan(\frac{\mu_{y, i}}{\mu_{x, i}})$, respectively. In the simplest case where the detector is in the molecular $z$ direction, as in Figure \ref{fig:effective_dipoles}a), we have that the projection onto the polarization-filtered component of the dipole is given by (dropping the site index $m$ for ease of notation)
\begin{align}
 \mu_{eff}(\phi) & = P(\phi)\vec{\mu} \nonumber \\ & = 
\begin{pmatrix}
\cos(\phi) & \sin(\phi)& 0
\end{pmatrix} 
\begin{pmatrix}
\mu_x \\
\mu_y \\
\mu_z
\end{pmatrix} \nonumber \\ &
= \mu [\cos(\phi) \sin(\eta) \cos(\theta) + \sin(\phi)\sin(\eta)\sin(\theta)] \nonumber \\&
= \mu \sin(\eta) \cos(\theta - \phi).
\end{align}
We note that for $N$ sites, there is thus a list of $N$ dipole moments each characterised by the above magnitude and relevant angles in the molecule frame. The effective polarization-filtered dipole operator is then given by $\hat{\mu}(\phi) = \sum_i^N \mu_i(\phi) |0\rangle \langle i| + H.c$ in the single excitation manifold.

We can also define different molecule-detector orientations via the angles $\theta_{det}$ and $\eta_{det}$ between the detector and molecular frames, as in Figure \ref{fig:effective_dipoles}. For finite $\theta_{det}$ in this case we may simply make the substitution $\phi \to \phi + \theta_{det}$. In Figure \ref{fig:effective_dipoles}b) we illustrate the case of $\eta_{det} = \frac{\pi}{2}, \, \theta_{det} = 0$, corresponding to the detector plane aligning with the molecular y axis. This such a scenario of finite $\eta_{det}$ we have a rotation of the molecular basis describing the dipole vector $\vec{\mu} = (\mu_x,\,\mu_y,\, \mu_z)$ with respect to the lab frame, via
    
\begin{align}
     &\vec{\mu}^\prime(\eta_{det}, 0) =  P(\phi)R(\eta_{det})\vec{\mu}\\&
    =
\begin{pmatrix}
\cos(\phi) & \sin(\phi)& 0
\end{pmatrix} 
\begin{pmatrix}
1 & 0 & 0 \\
0  & \cos(\eta_{det}) & \sin(\eta_{det})  \\
0 & -\sin(\eta_{det}) & \cos(\eta_{det}) 
\end{pmatrix} 
\begin{pmatrix}
\mu_x \\
\mu_y \\
\mu_z
\end{pmatrix}. \nonumber
\end{align}

Similarly, in \ref{fig:effective_dipoles}c) we have both finite $\eta_{det}$ and $\theta_{det}$, such that the relevant rotation becomes
\begin{widetext}

\begin{align}
    \vec{\mu}^\prime(\eta_{det}, \theta_{det}) & =  P(\phi)  R(\eta_{det}) S(\theta_{det}) \vec{\mu} \nonumber \\&
    =
\begin{pmatrix}
\cos(\phi) & \sin(\phi)& 0
\end{pmatrix} 
\begin{pmatrix}
1 & 0 & 0 \\
0  & \cos(\eta_{det}) & \sin(\eta_{det})  \\
0 & -\sin(\eta_{det}) & \cos(\eta_{det}) 
\end{pmatrix} 
\begin{pmatrix}
 \cos(\theta_{det}) & \sin(\theta_{det}) & 0  \\
-\sin(\theta_{det}) & \cos(\theta_{det})  & 0\\
0 & 0 & 1 
\end{pmatrix} 
\begin{pmatrix}
\mu_x \\
\mu_y \\
\mu_z
\end{pmatrix}.
\end{align}
\end{widetext}

In general, then, the effective dipole moment can be obtained by a two step process of the required rotations $\vec{\mu} \to \vec{\mu}^\prime$ such that the dipole aligns with the $z-$axis, followed by taking the projection in that basis $\vec{\mu}^\prime \to \mu^\prime_x \cos(\phi) + \mu^\prime_y\sin(\phi)$. 
We thus define the dipole operator for site $m$ as $\hat{\mu}_{\phi, m} = \vec{\mu}_{\phi}^\prime(\eta_{det}, \theta_{det}) |m\rangle \langle 0| + h.c$, and the total effective dipole operators as $\hat{\mu}_{\phi} = \sum_m \hat{\mu}_{\phi, m} $.

\end{widetext}
\renewcommand{\thefigure}{S\arabic{figure}}

\begin{figure}
    \includegraphics[width=0.49\textwidth]{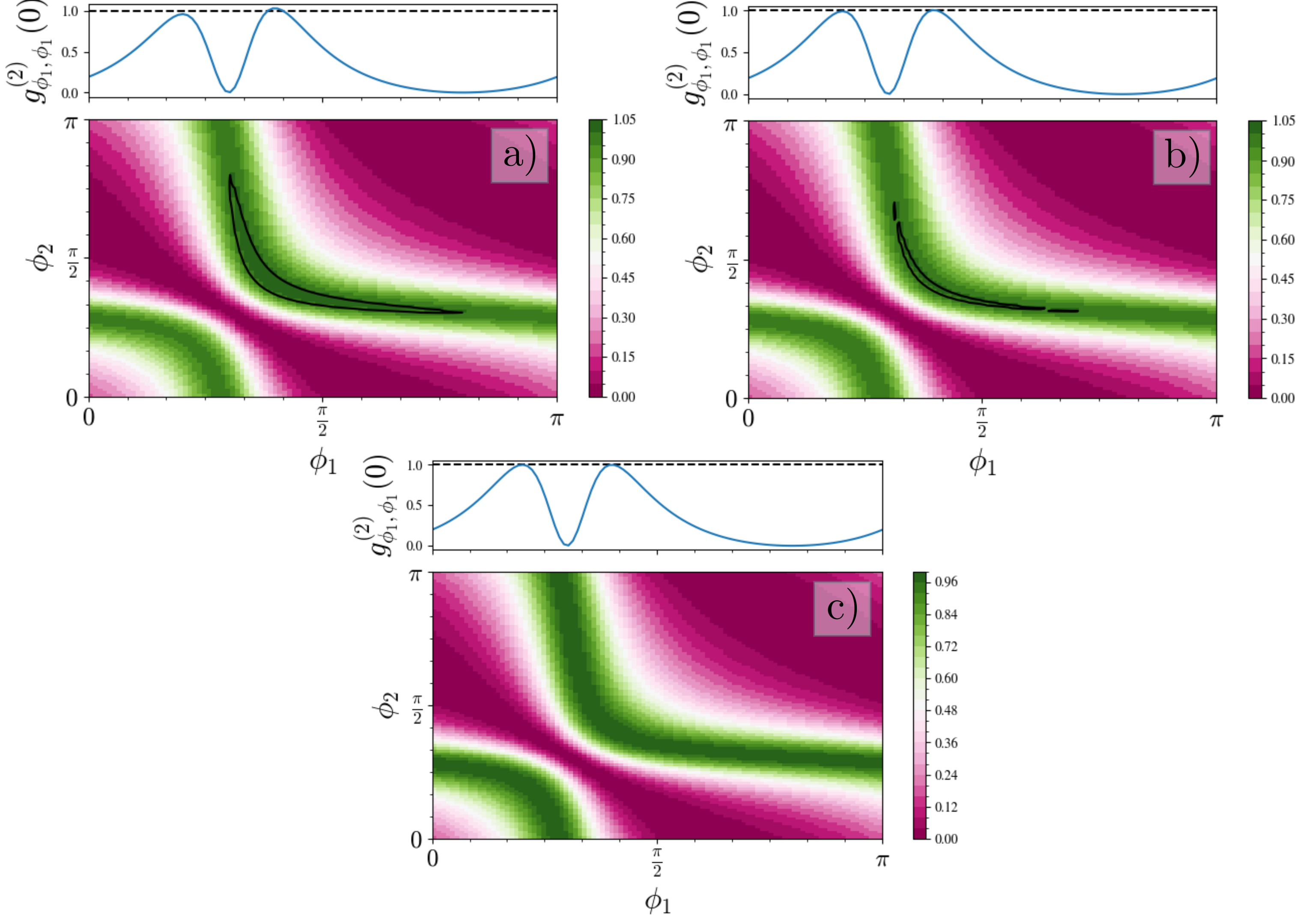}
    \caption{zero delay photon correlations for dimer model of Ref. \cite{SanchezMunoz2020} and the dependence on pure dephasing. a) Shows $\gamma_{pd} = 0$, b) shows $\gamma_{pd} = 10 \gamma$, c) shows $\gamma_{pd} = 1000 \gamma$. We note that the latter is closest to the case of photosynthetic systems, where the timescales of dephasing and decoherence due to vibrational background are $\sim 1000 \times$ faster (ps) than those of excited state decay to the EM environment (ns). }
    \label{fig:g20SMS}
\end{figure}

\begin{figure}
    \includegraphics[width=\linewidth]{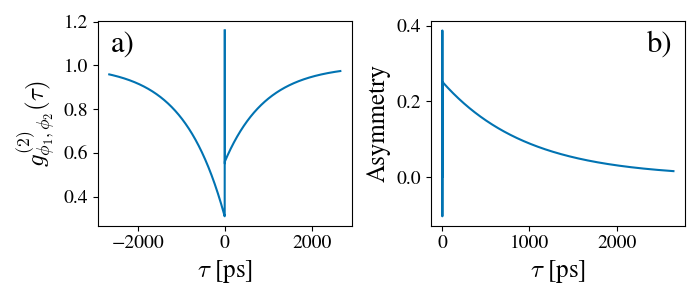}
        \includegraphics[width=\linewidth]{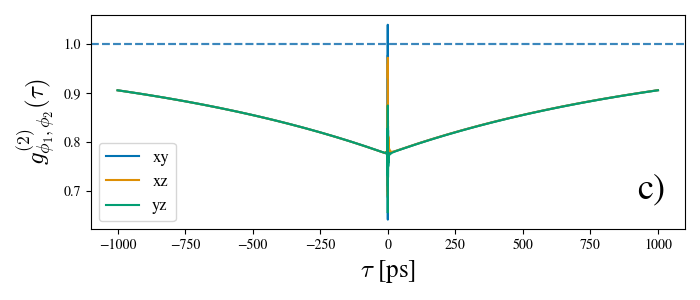}
    \caption{a) Long time dependence of photon cross-correlations for LH2 subunit, and b) the corresponding asymmetry. c) Shows long time dependence of photon cross-correlations of the FMO complex.}
    \label{fig:long_time}
\end{figure}

\begin{figure*}
    \includegraphics[width=\textwidth]{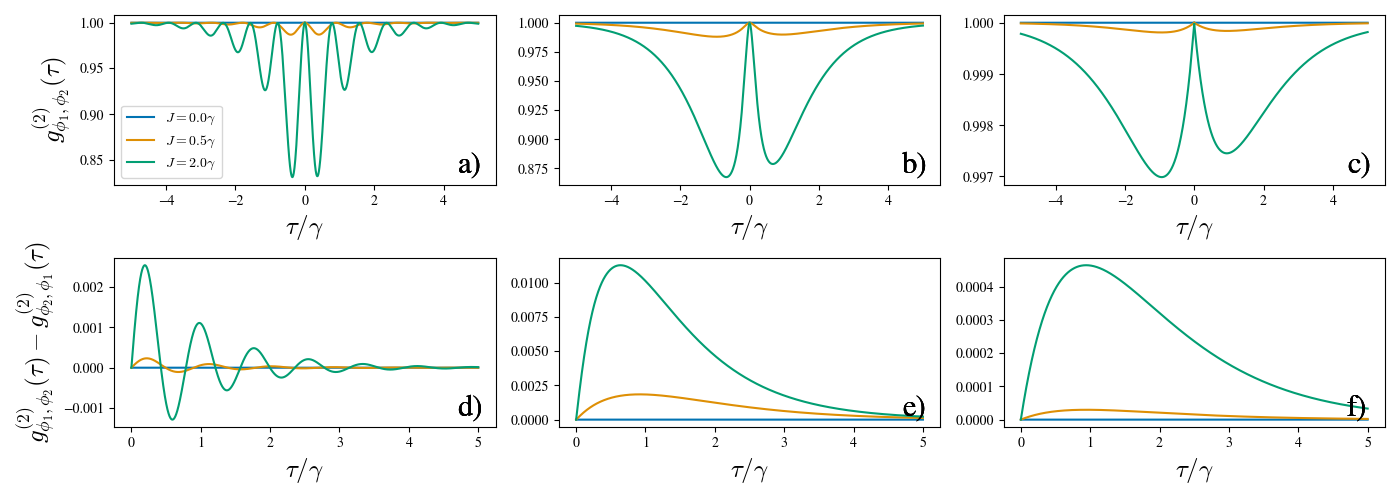}
    \caption{Correlation asymmetry dependence on coupling rate $J$ and pure dephasing rate $\gamma_{pd}$ for Dimer model. a), d) show $\gamma_{pd} = 0$ as in main text, b), e) show $\gamma_{pd} = 10 \gamma$, and c), f) show $\gamma_{pd} = 1000 \gamma$. For each plot $\phi_1 = 0, \, \phi_2 = \frac{\pi}{2}$.}
    \label{fig:dimer_g2t_gamma_pd}
\end{figure*}

\section{Additional numerical results}

In this section we provide some additional numerical analysis to supplement the conclusions of the main text. Firstly, for the Dimer model with identical parameters to that studied in Ref. \cite{SanchezMunoz2020} with an included Markovian environmental pure-dephasing, we show in Figure \ref{fig:g20SMS} that a Markovian environment can remove the ability of zero delay correlations to witness coherence in the steady-state for large values of pure dephasing $\gamma_{pd}$.

\begin{figure*}
    \includegraphics[width=\textwidth]{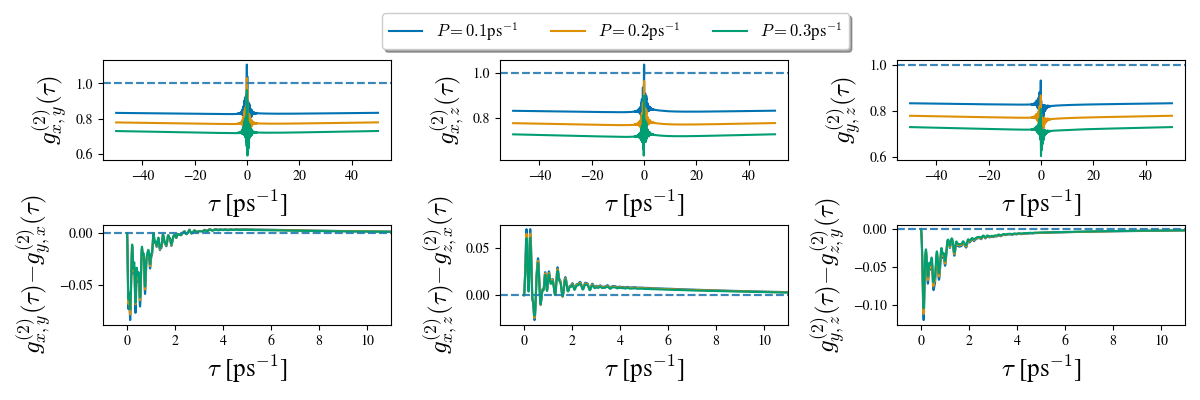}
    \caption{Pump power dependence of polarization filtered photon correlations a)-c) and their asymmetry d)-f) for the FMO complex. We use polarization filters corresponding to the $x, y,$ and $z$ components for simplicity, observing asymmetries in each case. We see that asymmetries can be observed over a few picoseconds, which is the typical timescale for exciton transport.}
    \label{fig:g2t_fmo}
\end{figure*}

In Figure \ref{fig:long_time} we show the time dependence of photon correlations for the LH2 subunit model and the FMO complex, showing indeed that at timescales longer than the $\sim 1 ns$ decay timescale of the electronic excited states photons correlations tend to the uncorrelated limit of $g^{(2)}_{\phi_1 \phi_2} = 1$. It is further notable that the correlation asymmetry similarly lasts for this long time, which is not observed in the FMO complex (see Figure \ref{fig:g2t_fmo}). This is due to the larger system size $M$ of FMO, combined with the essentially random projection of the polarization filter onto the exciton basis, leading to decaying asymmetry over the picosecond timescales of the vibrational environment. In the LH2 subunit, after this picosecond timescale some asymmetry remains as the projection of the polarization filter onto the excitonic basis biases strongly particular states. Future work will analyze in more detail the role of the projection of the polarization filter onto particular states onto the correlation asymmetry, which dictates the role of the vibrational environment in the decay of asymmetries over short times.

In Figure \ref{fig:dimer_g2t_gamma_pd} we similarly show how photon correlation asymmetry depends on the pure dephasing rate $\gamma_{pd}$. We observe that coherent oscillations are rapidly destroyed by the pure dephasing, and the observed asymmetries are significantly reduced. Nonetheless, for each case in which coherences in the steady-state are present, that is, for finite $V$, asymmetries in photon correlations are similarly present. In the case of $V = 0$ the dynamics is effectively completely classical, with no coherences in the steady-state, asymmetries are not observed in this case.

\begin{figure}
    \includegraphics[width=0.5\textwidth]{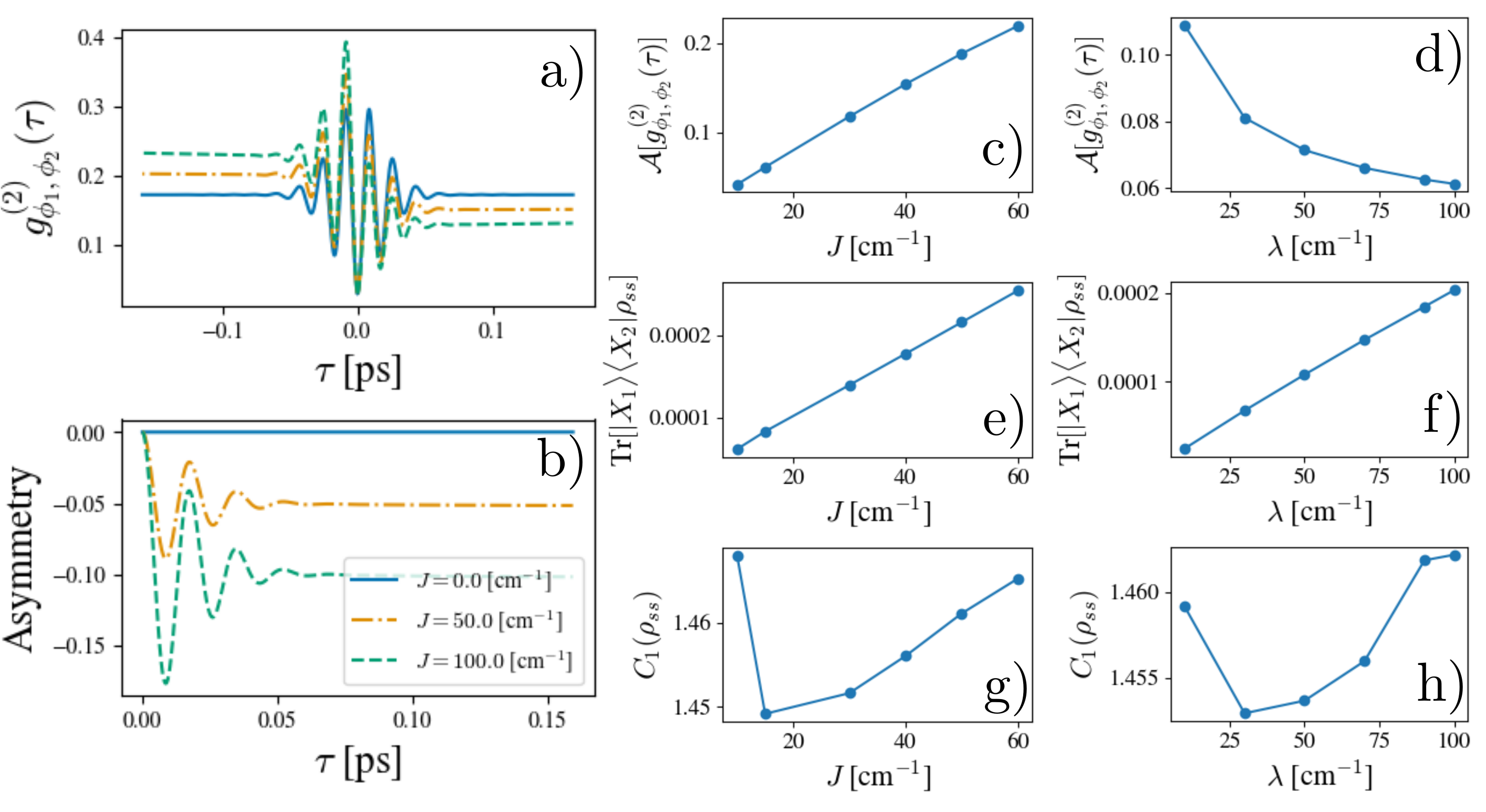}
    \caption{Photon correlations of the dimer model. Time dependence a), time-asymmetry b), total asymmetry c), d) over $\tau \approx 13.25$ ps $>> \tau_{S-E}$, excitonic coherence e), f) and basis independent coherence g), h) and their dependence on reorganization energy $\lambda$ and coupling strength $J$. $N_{trunc} = 58$ (note that this is a very large value as the basis independent coherence for the dimer model converges very slowly due to its high sensitivity to small changes in populations and coherences in the steady-state.)}
    \label{fig:dimer_correlations}
\end{figure}

In Figure \ref{fig:dimer_correlations} we show the PC645 inspired dimer model cross-correlation time-dependence, as well as the total asymmetries, excitonic coherences, and basis independent coherence, and how each change with both electronic coupling $J$, and reorganization energy $\lambda$. Here we see similar results to the LH2 subunit shown in Figure \ref{fig:trimer} of the main text, where the correlation asymmetry closely resembles the coherence measures dependence on electronic coupling, but deviates in its behaviour for reorganization energy, as whilst the steady-state indeed has higher excitonic and basis independent coherences as a result of the environmental coupling, the dynamics is more dominated by environmental degrees of freedom and is thus more incoherent. 

In Figure \ref{fig:dimer_correlations}a), b) we show the photon correlations and their asymmetry for the dimer model with $\lambda = \Omega = 100$ cm$^{-1}$, for various values of the inter-site electronic coupling $J$. We see that electronic coupling induces an asymmetry in photon correlations. In the $J=0$ case, the excitonic and site bases are identical, and the environment thus does not enable exciton transport or sustain coherences in the steady-state, leading to symmetric correlations upon interchange of measurement order. We see that asymmetries are introduced even for very weak inter-site couplings, and hence asymmetry acts as a very sensitive witness of coherent excitation transfer.

In the dimer model we see a more complex dependence of the basis independent coherence on reorganization energy, which can be explained as the singly excited state, which dominates the change in the basis independent coherence, only has two possible excitonic states. At weak pumping, there is no transfer, and thus pumping the highest energy exciton leads to a steady state with only population in this excitonic state, and thus a high $C_1(\rho)$ (low entropy), as $\lambda$ is increased, exciton transfer is enabled, reducing the excitonic coherence. For higher $\lambda$ values, however, transport to the lowest energy exciton is stronger than pumping of the high energy, and thus steady-state population is dominant in the low energy exciton, again increasing the basis independent coherence due to the low entropy of the state. Thus, we see that for the dimer model, $C_1(\rho)$ does not fully capture the relevant coherent properties of excitation transport, which decrease as the environment dominates this process. This is observed for the correlation asymmetry, as the transport is more incoherent as $\lambda$ is increased.

In Figure \ref{fig:g2t_fmo} we show how photon correlations and their asymmetries depend on the pump power in the FMO complex. We note that a change is observed in the values of $g^{(2)}$ as the pump power is altered for all delay times. This is due to the fact that the dynamical generator driving dynamics at early times is dominated by the inter-exciton dynamics, and independent of pump power, however the steady-state populations depend strongly on the pump power. Thus the time evolution follows near identical trends, with differing absolute values of the photon correlations.  The data in Figure \ref{fig:g2t_fmo} are examples of those used in the main text to obtain Figure \ref{fig:g2t}g), in which we observe a steady decrease in the asymmetry. 

\begin{figure}
    \includegraphics[width=0.45\textwidth]{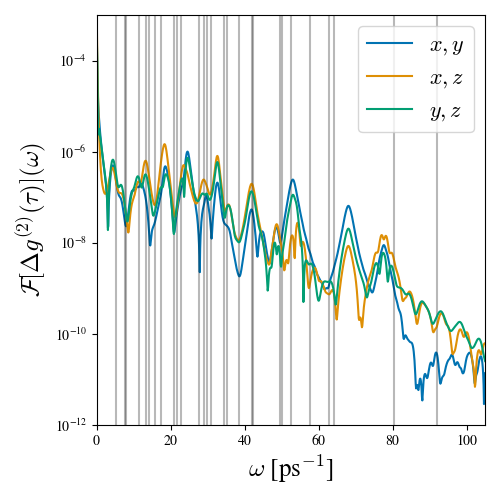}
    \caption{Fourier transform of photon correlation asymmetry of FMO. Solid vertical lines are all exciton energy gaps. We observe, indeed, that the prominent oscillation frequencies occur in the same frequency range as that expected in coherent inter-exciton transport.}
    \label{fig:asym_FT}
\end{figure}

We additionally in Figure \ref{fig:asym_FT} show the Fourier transform (exploiting the standard Blackman window filtering method) of the photon correlation asymmetry results in Figure \ref{fig:g2t}f) of the main text. Here we observe that, indeed, the prominent early time oscillatory behaviours are of comparable frequency to the exciton energy gaps. 

\begin{figure}
    \includegraphics[width=0.45\textwidth]{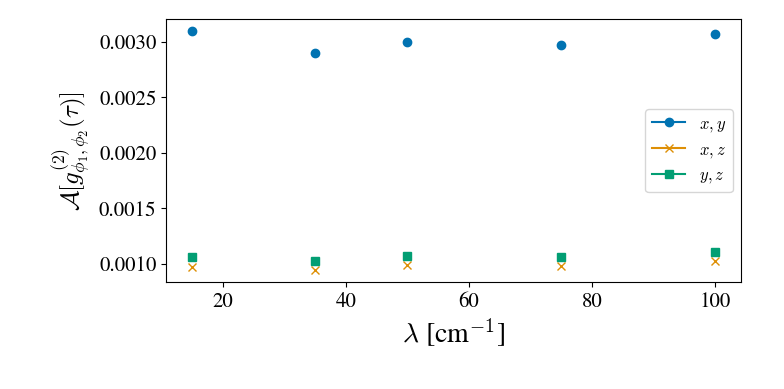}
    \caption{Asymmetry of FMO at early times, with integration time of asymmetry = 30 fs.}
    \label{fig:early_time_asym}
\end{figure}

In Figure \ref{fig:early_time_asym} we show the asymmetry of polarization filtered correlation functions for the FMO complex at early times, with the time integration of asymmetry taken as 30 fs. We see that as the reorganization energy $\lambda$ is increased the asymmetry for some values indeed increases, which manifests as a result of early time environmentally induced coherent transfer. This effect is much smaller in FMO compared to the LH2 subunit shown in the main text, which we associate to the multichromophoric system having many contributions to the transport in different directions, leading to a washing out of the induced asymmetries.

\begin{figure}
    \includegraphics[width=0.45\textwidth]{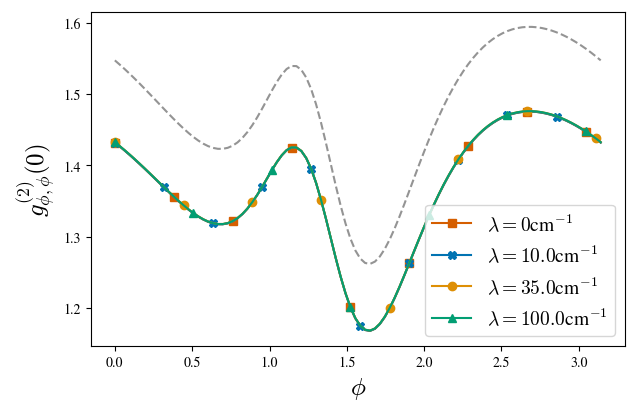}
    \caption{Zero time polarization filtered correlations for varying reorganization energy (solid line), and their comparison to the tightened classical bound (dashed line). We see here that the zero-delay correlation functions change very slowly with environmental parameters of the background Drude-Lorentz mode. There is no violation of the zero-delay bound for any value of the reorganization energy.}
    \label{fig:better_bound}
\end{figure}

Finally, in Figure \ref{fig:better_bound} we show that zero time photon correlations are robust in their form to environmental changes in FMO by altering the Drude-Lorentz reorganisation energy, observing only very slight changes in the zero-delay correlations over a wide range of environmental parameters. We note that this is consistent with the results observed for the dimer model, where whilst we observe that the region of violation of the zero-delay bound indeed changes, this is around regions where a very small change in the zero-delay values has a large effect. We further show that a tightened form of the bound in Ref. \cite{SanchezMunoz2020}, written as
\begin{align}
    g^{(2)}_{\phi_1, \phi_1}(0) \leq 2 - \frac{2 \sum_\alpha |\mu_{\phi_1, \alpha}|^2 \langle \sigma_\alpha^\dagger \sigma_\alpha \rangle }{\sum_{\alpha, \beta} |\mu_{\phi_1, \alpha}|^2 |\mu_{\phi_1, \beta}|^2 \langle \sigma_\alpha^\dagger \sigma_\alpha \rangle \langle \sigma_\beta^\dagger \sigma_\beta \rangle },
\end{align}
where $\mu_{\phi_1, \alpha}$ is the component of the dipole moment on exciton $\alpha$, and $\sigma_\alpha = |0\rangle \langle \alpha|$, is also not violated for any of the shown angles in FMO.

\section{Combined Nakajima-Zwanzig Hierarchical equations of motion approach }

\subsection{Set up}

Here we apply the Nakajima-Zwanzig (NZ) formalism to the hierarchical equations of motion (HEOM) in order to obtain a non-additive description of both the interaction with the non-Markovian environment of the pigment protein complex via HEOM, as well as the incoherent Markovian interaction with the electromagnetic environment and pump. We begin by writing our Hamiltonian in the form
\begin{equation}
H = H_S + H_{E_0} + H_{SE_0} + \sum_{n=1}^{N_{heom}}(H_{E_n} + H_{SE_n}),     
\end{equation}
where our goal is to treat the environmental contribution of $H_{E_0}$ perturbatively, and the other $N_{heom}$ environmental contributions with HEOM. We thus redefine $H_{S^\prime} = H_S + H_{E_0} + H_{SE_0}$, which in our case is the molecular electronic degrees of freedom $H_S$, and the perturbative environment $E_0$ of the pump and decay into the electromagnetic field. We indicate a density operator of the combined space of $S$ and $E_0$ as $\sigma$. We write the interaction Hamiltonian between the molecular electronic degrees of freedom and its vibrational environment, which we wish to treat non-perturbatively via HEOM, as
\begin{align}
    H_{E_n} = V_n \otimes B_n,
\end{align}
where $V_n$ is a system operator on the space of $H_S$, and $B_n$ a bath operator on that of $H_{B, n}$.

HEOM is a formalism for non-perturbative simulations of open systems with exponentially decaying bath correlation functions $C_m(t)$ on each site $m$ \cite{ishizaki2006, ishizaki2009},
\begin{equation}\label{eq:bath_corr_func}
    C_m(t) = \sum_k^{K_m} c_{m, k} e^{-\gamma_{m, k} t}.
\end{equation}
We then have a set of coefficients $c_{m,k}$, as well as a corresponding set of rates $\gamma_{m, k}$, and operators $B_{m, k} = B_m$. We relabel $k$ by merging these indices into a single index $k \in [1, K]$, with $K = \sum_{m}K_m$. The formalism then becomes identical to the case of a single bath, except that each of the coefficients may now have different coupling operators.

Then the HEOM describe a set of coupled equations of motion for a set of auxiliary density operators (ADOs) $\sigma_{\mathbf{n}}$, where the multi-index $\mathbf{n} = (n_1, \cdots, n_K)$ of the ADO is a set of $K$ positive integers, where $\sum_k^K \mathbf{n}_k = T_{\mathbf{n}}$ is the tier of the ADO. The zero tier consists of one ADO only, which is the reduced density operator of the system, and the coupling additional ADOs captures non-perturbative dynamics of the environment. The ADOs are coupled between adjacent tiers, and not within the same tier. 

For the Drude-Lorentz environments treated here we have
\begin{align}
    c_0 & = \lambda\Omega\left(\cot\left(\frac{\beta\Omega}{2}\right) - i\right), \\
    c_k & = \frac{4\lambda\Omega}{\beta}\frac{\nu_k}{\nu_k^2 - \Omega^2}, \\
    \gamma_0 & = \Omega, \\
    \gamma_k & = \nu_k,
\end{align}
with $k = 1, 2, 3\ldots$ and $\nu_k = \frac{2\pi k}{\beta}$, with $\beta = (k_B T)^{-1}$ is the inverse temperature, with $k_B = 1$ the Boltzmann constant. $\nu_k$ are the Matsubara frequencies. We note that for high temperatures treated here the expansion can be cut-off at simply the zero order term $c_0, \gamma_0$.

The HEOM for $H_{S^\prime}$ can be written as \cite{ishizaki2009}
\begin{align}
   \partial_t \sigma_{\mathbf{n}}(t) & = \mathfrak{L}^D_{\mathbf{n}}\sigma_{\mathbf{n}}(t)  + \sum_k^K ( \mathfrak{L}^+_{\mathbf{n}}\sigma_{\mathbf{n}_{k+
}}(t) +  \mathfrak{L}^{-}_{\mathbf{n}}\sigma_{\mathbf{n}_{k-}}(t))\nonumber \\& 
= \sum_{\mathbf{n}}\mathfrak{L}_{\mathbf{n}}\sigma_{\mathbf{n}^\prime}(t) \\ & \nonumber 
= \sum_{\mathbf{n}^\prime \in \{\mathbf{n}+{k-}, \mathbf{n}, \mathbf{n}_{k+} \}}\mathfrak{L}_{\mathbf{n}}\sigma_{\mathbf{n}^\prime}(t),
\end{align}
with $\mathfrak{L}^D_{\mathbf{n}}[\sigma_{\mathbf{n}}(t)] = \mathfrak{L}_0[\sigma_{\mathbf{n}}(t)] - \sum_{n_k \in \mathbf{n} } n_k \nu_k \mathbb{1}_{E_0}$ describing the `diagonal' contribution of the ADO, where $\mathfrak{L}_0 = -iH_{S^\prime}^\times$, and the inter-tier couplings are $\mathfrak{L}_{m,k}^- = -in_{m,k}(\text{Re}(c_{m,k})V_m^\times + i\text{Im}(c_{m,k})V_m^\circ)$, $\mathfrak{L}_{m,k}^+ = -iV_m^\times$, and $\mathbf{n}_{k\pm}$ differs from $\mathbf{n}$ by $\pm 1$ in the $(k)$-th entry of the tuple $\mathbf{n}$, $n_{k}$.  Here we have used the notation $O^\times \cdot := [O, \cdot]$ and $O^\circ \cdot := [O, \cdot]_+$.

Note that the hierarchical expansion that describes the environments $E_i \, |\, i \geq 1$ has not altered the form of the environment $E_0$ or it's coupling to the system, however, the effect of this latter environment alters the self interaction term $\mathfrak{L}_{\mathbf{n}}^D(t)$ at every level of the hierarchy. We note that in our calculations we exploit the Ishizaki-Tanimura truncation scheme \cite{Ishizaki2005Dec}, which adds a term to the diagonal $\mathfrak{L}^D_{\mathbf{n}}$ aiding in convergence of the Hierarchy, which however does not alter otherwise the form of the HEOM. 

Crucially, $E_0$ also effects the form of each ADO, as $\rho^\prime_{\mathbf{n}}(0) = \rho_{\mathbf{n}, S}(0) \otimes \rho_{\mathbf{n}, E_0}(0)$. If we wish to trace out the environment $E_0$, we must thus do so at every tier in the hierarchy.

We note that in the absence of the environment $E_0$, we have the regular HEOM acting on density operators $\rho \in \mathcal{H}$, which is similarly written as
\begin{align}
   \partial_t \rho_{\mathbf{n}}(t) & = \mathcal{L}^D_{\mathbf{n}}\rho_{\mathbf{n}}(t)  + \sum_k^K ( \mathcal{L}^+_{\mathbf{n}}\rho_{\mathbf{n}_{k+
}}(t) +  \mathcal{L}^{-}_{\mathbf{n}}\rho_{\mathbf{n}_{k-}}(t)).
\end{align}
with $\mathcal{L}^D_{\mathbf{n}}[\rho_{\mathbf{n}}(t)] = -i [H_S, \rho_{\mathbf{n}}(t)] - \sum_{n_k \in \mathbf{n} } n_k \nu_k \mathbb{1}_{E_0}$, and similarly to above,  $\mathcal{L}_0 = -iH_S^\times$, $\mathcal{L}_{m,k}^- = -in_{m,k}(\text{Re}(c_{m,k})V_m^\times + i\text{Im}(c_{m,k})V_m^\circ)$, $\mathcal{L}_{m,k}^+ = -iV_m^\times$. Similarly, if only the environment $E_0$ was present, and we treated this as a Markovian environment of GKSL form, we would write
\begin{equation}
    \partial_t \rho_{\mathbf{0}}(t) = \mathcal{L_S}\rho_{\mathbf{0}}(t) + \mathcal{D}_{E_0}\rho_{\mathbf{0}}(t)
\end{equation}
where $\mathcal{D}_{E_0}$ is the GKSL dissipator. Were we to treat these two environments under an additive approximation, we could simply replace $\mathcal{L}^D_{\mathbf{0}} \to \mathcal{L}^D_{\mathbf{0}} + \mathcal{D}_{E_0}$, where the additional term acts only on the system density operator $\rho_\mathbf{0}$. We will see, however, that the dissipation must act on every tier of the hierarchy.

\subsection{Derivation of NZ-HEOM}

We begin by writing total density operator of the entire HEOM space as the total density operator as $\sigma_{\Delta}(t) = \sum_{\mathbf{n}} \sigma_{\mathbf{n}}$, with 
\begin{align}\label{eq:HEOM_liouvillian}
    \partial_t \sigma_{\Delta}(t) = \mathfrak{L}_{\Delta}[\sigma_{\Delta} (t)],
\end{align}
in the following we use a subscripted ${\Delta}$ to refer to the entire hierarchy of coupled ADOs - the space of these ADOs, either $\mathcal{H}_S$ or $\mathcal{H}_S \otimes \mathcal{H}_{E_0}$, is denoted by the notation $\mathcal{L}, \rho_{\mathbf{n}}$ or $\mathfrak{L}, \sigma_{\mathbf{n}}$, respectively. Our approach will be similar to that in Ref. \cite{fay2022}, involving tracing out the $E_0$ environment with a projection operator approach to obtain a hierarchical set of equations for $\rho_{\mathbf{n}}$ from the exact HEOM of Eq. \eqref{eq:HEOM_liouvillian}. 

There are multiple formulations of projection operator based approaches to the description of open quantum systems, the most well known of which are Nakajima-Zwanzig \cite{nakajima, zwanzig1960} (NZ) and time-convolutionless (TCL) \cite{tokuyama1976} master equations. Projection operator techniques rely on the definition of projectors $\mathcal{P}$ onto the `relevant' Hilbert space that defines the open system, and it's complement $\mathcal{Q}$, the `irrelevant' part. Here we follow the NZ approach, which is summarized in its more standard form in Section \ref{app:NZ} below.

The total system Liouvillian can be written as $\mathfrak{L}_0 = \mathfrak{L}_S + \mathfrak{L}_{E_0} + \alpha \mathfrak{L}_{SE_0}$, where we introduce the unitless coupling strength $\alpha$ for later convenience in keeping track of the order in the interaction between the system part $S$ and perturbative environment $E_0$. 
We then define the projection operators $\mathcal{P}_{\mathbf{n}}$ by $\mathcal{P}_{\mathbf{n}} \sigma_{\Delta} = \Tr_{E_0}[\sigma_{\mathbf{n}}] \otimes \rho_{E_0}^{\beta} = \rho_{\mathbf{n}} \otimes \rho_{E_0}^{\beta}$, where $\rho_{E_0}^{\beta}$ is the thermal state of $E_0$ at inverse temperature $\beta$ (note that in general this can be any reference state, typically chosen to be a Gaussian state). The irrelevant part projector is then $\mathcal{Q} = \mathbb{1} - \mathcal{P} = \mathbb{1} - \sum_{\mathbf{n}} \mathcal{P}_{\mathbf{n}} =  \sum_{\mathbf{n}}(\mathbb{1}_{\mathbf{n}} -  \mathcal{P}_{\mathbf{n}}) = \sum_{\mathbf{n}}\mathcal{Q}_{\mathbf{n}}$.

We can now follow the standard NZ projection operator technique, however in this case, there are many `relevant' projection operators that together form a coupled set of equations
\begin{align}\label{eq:relevant}
    \partial_t \mathcal{P}_{\mathbf{n}} \sigma_{\Delta} (t) &= \mathcal{P}_{\mathbf{n}}  \mathfrak{L}_{\Delta} \sigma_{\Delta} (t) \\ & 
    = \sum_{\mathbf{n}^\prime} \mathcal{P}_{\mathbf{n}} \mathfrak{L}_{\Delta} \mathcal{P}_{\mathbf{n}^\prime} \sigma_{\Delta} (t) + \mathcal{P}_{\mathbf{n}} \mathfrak{L}_{\Delta} \mathcal{Q}\sigma_{\Delta} (t) \nonumber
\end{align}
and for the irrelevant part
\begin{align}
    \partial_t \mathcal{Q}\sigma_{\Delta} (t) &= \mathcal{Q} \mathfrak{L}_{\Delta}\sigma_{\Delta} (t) \\ & 
    = \sum_{\mathbf{n}^\prime} \mathcal{Q} \mathfrak{L}_{\Delta} \mathcal{P}_{\mathbf{n}^\prime} \sigma_{\Delta} (t) + \mathcal{Q} \mathfrak{L}_{\Delta} \mathcal{Q} \sigma_{\Delta} (t) \nonumber.
\end{align}

We can then take the Laplace transform of each of the above two expressions to find
\begin{align}
    s\mathcal{P}_{\mathbf{n}} \tilde{\sigma}_{\Delta} (s) & - \mathcal{P}_{\mathbf{n}} \sigma_{\Delta} (0)  \\ & 
    = \sum_{\mathbf{n}^\prime} \mathcal{P}_{\mathbf{n}} \mathfrak{L}_{\Delta} \mathcal{P}_{\mathbf{n}^\prime} \tilde{\sigma}_{\Delta} (s)  + \mathcal{P}_{\mathbf{n}} \mathfrak{L}_{\Delta} \mathcal{Q}\tilde{\sigma}_{\Delta} (s)  \nonumber
\end{align}
and
\begin{align}\label{eq:irrelevant}
    s\mathcal{Q}\tilde{\sigma}_{\Delta} (s) & - \mathcal{Q} \sigma_{\Delta} (0) \\ & 
    =  \sum_{\mathbf{n}^\prime} \mathcal{Q} \mathfrak{L}_{\Delta} \mathcal{P}_{\mathbf{n}^\prime} \tilde{\sigma}_{\Delta} (s)  + \mathcal{Q} \mathfrak{L}_{\Delta} \mathcal{Q}\tilde{\sigma}_{\Delta} (s)  \nonumber
\end{align}
where $\tilde{\sigma}(s)_{\Delta} = (s - \mathfrak{L}_{\Delta})\sigma(0)$. We can rearrange Eq. \eqref{eq:irrelevant} to obtain
\begin{align}
    \mathcal{Q}\tilde{\sigma}_{\Delta} (s) = \frac{1}{s - \mathcal{Q} \mathfrak{L}_\Delta } (\mathcal{Q} \sigma_{\Delta} (0) +  \sum_{\mathbf{n}^\prime} \mathcal{Q} \mathfrak{L}_{\Delta} \mathcal{P}_{\mathbf{n}^\prime} \tilde{\sigma}_{\Delta} (s))
\end{align}
and then substitute this expression for the irrelevant part into that for the relevant part, obtaining
\begin{align}
    s\mathcal{P}_{\mathbf{n}} & \tilde{\sigma}_{\Delta} (s) - \mathcal{P}_{\mathbf{n}} \sigma_{\Delta} (0) \nonumber \\ & 
    = \sum_{\mathbf{n}^\prime} \mathcal{P}_{\mathbf{n}} \mathfrak{L}_{\Delta} \mathcal{P}_{\mathbf{n}^\prime} \tilde{\sigma}_{\Delta} (s) \\ &  + \mathcal{P}_{\mathbf{n}} \mathfrak{L}_{\Delta} \frac{1}{s - \mathcal{Q} \mathfrak{L}_\Delta } (\mathcal{Q} \sigma_{\Delta} (0) +  \sum_{\mathbf{n}^\prime} \mathcal{Q} \mathfrak{L}_{\Delta} \mathcal{P}_{\mathbf{n}^\prime} \tilde{\sigma}_{\Delta} (s)) . \nonumber
\end{align}
The inverse Laplace transform then gives us
\begin{align}
    \partial_t \mathcal{P}_{\mathbf{n}} \sigma_{\Delta} (t) =  \sum_{\mathbf{n}^\prime}& \mathcal{P}_{\mathbf{n}} \mathfrak{L}_{\Delta} \mathcal{P}_{\mathbf{n}^\prime} \sigma_{\Delta} (t) + \mathcal{P}_{\mathbf{n}} \mathfrak{L}_{\Delta} e^{\mathcal{Q} \mathfrak{L}_{\Delta} t} \mathcal{Q} \sigma_{\Delta}(0) \nonumber \\ &
    +\sum_{\mathbf{n}^\prime} \int_0^t d\tau \mathcal{K}_{\mathbf{n}, \mathbf{n}^\prime}(t - \tau) \sigma_{\Delta}(\tau),
\end{align}
where we have defined the superoperator valued NZ kernel $\mathcal{K}_{\mathbf{n}, \mathbf{n}^\prime}(t) = \mathcal{P}_{\mathbf{n}} \mathfrak{L}_{\Delta} e^{\mathcal{Q} \mathfrak{L}_{\Delta} t} \mathcal{Q}  \mathfrak{L}_{\Delta} \mathcal{P}_{\mathbf{n}^\prime} $. We thus see that the NZ kernel may, in general, mix tiers of the ADOs. In the following we ignore the second term on the RHS of the above expression, the so-called inhomogeneous term, which identically vanishes for product state initial states.

\subsection{Weak coupling limit: Markov NZ-HEOM}

The set of projections onto $\mathfrak{L}_{\Delta}$ define the elements of the dynamical generator in the relevant subspace:
\begin{align}
   \mathcal{P}_{\mathbf{n}} \mathfrak{L}_{\Delta} &  \mathcal{P}_{\mathbf{n}^\prime}= \\ & \mathcal{P}_{\mathbf{n}}  (\mathfrak{L}_\mathbf{n}^D {\delta}_{\mathbf{n} \mathbf{n}^\prime} + \sum_k^K ( \mathfrak{L}^+_{\mathbf{n}} {\delta}_{\mathbf{n} \mathbf{n}^\prime_{k+}} +  \mathfrak{L}^{-}_{\mathbf{n}} {\delta}_{\mathbf{n} \mathbf{n}^\prime_{k-}}  )\mathcal{P}_{\mathbf{n}^\prime}\nonumber 
\end{align}
we then have projectors mixing relevant and irrelevant subspaces,
\begin{subequations}\label{eq:heom_projection_permutations}
    \begin{eqnarray}
          &\mathcal{P}_{\mathbf{n}}\mathfrak{L}_{\Delta} \mathcal{Q} = \alpha \mathcal{P}_{\mathbf{n}}\mathfrak{L}_{SE_0} \mathcal{Q}\\&
          \mathcal{Q}\mathfrak{L}_{\Delta} \mathcal{P}_{\mathbf{n}} = \alpha \mathcal{Q}\mathfrak{L}_{SE_0} \mathcal{P}_{\mathbf{n}}
    \end{eqnarray}
\end{subequations}
and finally the irrelevant subspace contributions, which we can see differ for each ADO. Using that $\mathcal{Q} = \mathbb{1} - \sum_{\mathbf{n}} \mathcal{P}_{\mathbf{n}} = \sum_{\mathbf{n}} (\mathbb{1}_{\mathbf{n}} - \mathcal{P}_{\mathbf{n}})$, we can define $\mathcal{Q}_{\mathbf{n}} = \mathbb{1}_{\mathbf{n}} - \mathcal{P}_{\mathbf{n}}$, and write
\begin{align}\label{eq:heom_QLQ}
        \mathcal{Q}\mathfrak{L}_{\Delta} \mathcal{Q} & = \sum_{\mathbf{n}, \mathbf{n}^\prime} \mathcal{Q}_{\mathbf{n}} \mathfrak{L}_{\Delta} \mathcal{Q}_{\mathbf{n}}  \nonumber\\ &
        = \sum_{\mathbf{n}} \mathcal{Q}_{\mathbf{n}} \mathfrak{L}_{\mathbf{n}} \mathcal{Q}_{\mathbf{n}} \nonumber \\&
        =  \sum_{\mathbf{n}} (\mathbb{1}_{\mathbf{n}} - \mathcal{P}_{\mathbf{n}}) \mathfrak{L}_{\mathbf{n}} \mathcal{Q}_{\mathbf{n}}  \\ & \nonumber 
        =  \sum_{\mathbf{n}} \mathfrak{L}_{\mathbf{n}} \mathcal{Q}_{\mathbf{n}} - \mathcal{P}_{\mathbf{n}} \mathfrak{L}_{SE_0, \mathbf{n}} \mathcal{Q}_{\mathbf{n}}  \\ & \nonumber 
        = \sum_{\mathbf{n}}   \mathfrak{L}_{\mathbf{n}} \mathcal{Q}_{\mathbf{n}}  - (\mathbb{1}_{\mathbf{n}} - \mathcal{Q}_{\mathbf{n}}) \mathfrak{L}_{SE_0, \mathbf{n}} \mathcal{Q}_{\mathbf{n}} \nonumber \\&
         = \sum_{\mathbf{n}} (\mathfrak{L}_{S, \mathbf{n}}\mathcal{Q} + \mathfrak{L}_{E_0, \mathbf{n}}\mathcal{Q} + \mathcal{Q}\mathfrak{L}_{SE_0, \mathbf{n}}\mathcal{Q}). \nonumber
\end{align}
where in the second line we have used that the irrelevant subspace, that including the interaction term with $E_0$, has no terms that mix tiers of the hierarchy, and in the final line defined 
\begin{equation}
    \mathfrak{L}_{S, \mathbf{n}} = \mathfrak{L}_S - \sum_{n_k \in \mathbf{n} } n_k \nu_k \mathbb{1}_S + \sum_k (\mathfrak{L}^+_{\mathbf{n}_{k+}} + \mathfrak{L}^-_{\mathbf{n}_{k-}}),
\end{equation}
and $\mathfrak{L}_{E_0, \mathbf{n}} = \mathfrak{L}_{E_0} - \sum_{n_k \in \mathbf{n} } n_k \nu_k \mathbb{1}_{E_0}$. We have included the subscript $\mathbf{n}$ in the $SE_0$ coupling to specify on which ADO they act, though note these terms are identical for each ADO. We then wish to obtain the HEOM including the influence of $H_{E_0}$ up to second-order in $\alpha$, and to do so, as above, take the Laplace transform of the relevant part(s) of the full ADOs given in Eq. \eqref{eq:relevant} 
\begin{widetext}
\begin{align}
          s\mathcal{P}_{\mathbf{n}}  &\tilde{\sigma}_{\Delta}(s)= \mathcal{P}_{\mathbf{n}} \tilde{\sigma}(0) + \sum_{\mathbf{n}^\prime}\mathcal{P}_{\mathbf{n}}\mathfrak{L}_{\Delta} \mathcal{P}_{\mathbf{n}^\prime}\tilde{\sigma}_{\Delta}(s)
          + \alpha^2 \mathcal{P}_{\mathbf{n}} \mathfrak{L}_{\Delta} \frac{1}{s - \mathcal{Q}\mathfrak{L}_{\Delta}}\mathfrak{L}_{\Delta} \sum_{\mathbf{n}^\prime} \mathcal{P}_{\mathbf{n}^\prime}\tilde{\sigma}_{\Delta}(s) .
\end{align}
We note that this equation is exact up to the inhomogeneous term, which is identically zero for an initial product state $\sigma_{\mathbf{n}}(0) = \rho_{\mathbf{n}}(0) \otimes \rho_{E_0}^\beta$ for each ADO. Using Eqs \eqref{eq:heom_projection_permutations} and \eqref{eq:heom_QLQ} we obtain
\begin{align}
          s\mathcal{P}_{\mathbf{n}}\tilde{\sigma}_{\Delta}(s) = \mathcal{P}_{\mathbf{n}}&  \tilde{\sigma}(0) + \sum_{\mathbf{n}^\prime}\mathcal{P}_{\mathbf{n}}\mathfrak{L}_{\Delta} \mathcal{P}_{\mathbf{n}^\prime}\tilde{\sigma}_{\Delta}(s) \\  & 
          + \alpha^2 \sum_{\mathbf{n}^\prime}\mathcal{P}_{\mathbf{n}} \mathfrak{L}_{SE_0, \mathbf{n}} \frac{1}{s - (\mathfrak{L}_{S, \mathbf{n}^{\prime}}\mathcal{Q}_{\mathbf{n}^{\prime}} + \mathfrak{L}_{E_0, \mathbf{n}^{\prime}}\mathcal{Q}_{\mathbf{n}^{\prime}} + \alpha\mathcal{Q}_{\mathbf{n}^{\prime}} \mathfrak{L}_{SE_0, \mathbf{n}^{\prime}}\mathcal{Q}_{\mathbf{n}^{\prime}})}  \mathfrak{L}_{SE_0, \mathbf{n}} \mathcal{P}_{\mathbf{n}^\prime}\tilde{\sigma}_{\Delta}(s) \nonumber.
\end{align}
We now treat the interaction with $E_0$ to lowest order in $\alpha$,
obtaining
\begin{align}
          s\mathcal{P}_{\mathbf{n}} \tilde{\sigma}_{\Delta}(s) = \mathcal{P}_{\mathbf{n}} \tilde{\sigma}(0) +& \sum_{\mathbf{n}^\prime}\mathcal{P}_{\mathbf{n}}\mathfrak{L}_{\Delta} \mathcal{P}_{\mathbf{n}^\prime}\tilde{\sigma}_{\Delta}(s)
          + \alpha^2\sum_{\mathbf{n}^\prime} \mathcal{P}_{\mathbf{n}} \mathfrak{L}_{SE_0} \frac{1}{s -  (\mathfrak{L}_{S, \mathbf{n}^{\prime}} + \mathfrak{L}_{E_0, \mathbf{n}^{\prime}})}\mathfrak{L}_{SE_0}  \mathcal{P}_{\mathbf{n}^\prime}\tilde{\sigma}_{\Delta}(s).
\end{align}
Performing the inverse Laplace transform we recover
\begin{align}
    \frac{d}{dt} \mathcal{P}_{\mathbf{n}} \sigma_{\Delta}(t) & = \sum_{\mathbf{n}^\prime} \mathcal{P}_{\mathbf{n}} \mathfrak{L}_{\Delta} \mathcal{P}_{\mathbf{n}^\prime} \sigma_{{\Delta}(t)} + \alpha^2 \sum_{\mathbf{n}^\prime}\int_0^t d\tau \mathcal{P}_{\mathbf{n}}\mathfrak{L}_{SE_0}  e^{(\mathfrak{L}_{S, \mathbf{n}^{\prime}} + \mathfrak{L}_{E_0, \mathbf{n}^{\prime}} )(t - \tau)} \mathfrak{L}_{SE_0}  \mathcal{P}_{\mathbf{n}^\prime} \sigma_{\Delta}(\tau) \\&
    = \sum_{\mathbf{n}^\prime} \Tr_{E_0, \mathbf{n}}[ \mathfrak{L}_{\Delta} \Tr_{E_0, \mathbf{n}^\prime}[ \sigma_{{\Delta}(t)} ] \otimes \rho_{E_0, \mathbf{n}^\prime}^\beta ]\otimes \rho_{E_0, \mathbf{n}}^\beta \nonumber \\& \qquad \qquad \qquad + \alpha^2 \sum_{\mathbf{n}^\prime}\int_0^t d\tau \Tr_{E_0, \mathbf{n}}[ \mathfrak{L}_{SE_0}  e^{(\mathfrak{L}_{S, \mathbf{n}^{\prime}} + \mathfrak{L}_{E_0, \mathbf{n}^{\prime}} )(t - \tau)} \mathfrak{L}_{SE_0}  \Tr_{E_0, \mathbf{n}^\prime}[  \sigma_{\Delta}(\tau) ] \otimes \rho_{E_0, \mathbf{n}^\prime}^\beta  ]\otimes \rho_{E_0, \mathbf{n}}^\beta\nonumber
\end{align}
where we have used that $\sum_{\mathbf{n}} \mathcal{P}_{\mathbf{n}} \sigma_{\Delta} = \sum_{\mathbf{n}} \Tr_{E_0, \mathbf{n}}[\sigma_{\mathbf{n}}] \otimes \rho_{E_0, \mathbf{n}} = \sum_{\mathbf{n}} \rho_\mathbf{n} \otimes \rho_{E_0, \mathbf{n}} $ to obtain the second-order NZ/HEOM equation for each reduced ADO by tracing over $E_0$
such that
\begin{align}\label{eq:NZ-HEOM_general}
    \frac{d}{dt} \rho_{\mathbf{n}}(t)  = \mathcal{L}^D_{\mathbf{n}}\rho_{\mathbf{n}}(t)  + \sum_k^K ( \mathcal{L}^+_{\mathbf{n}}\rho_{\mathbf{n}_{k+}}(t) +  \mathcal{L}^{-}_{\mathbf{n}}\rho_{\mathbf{n}_{K-}}(t))  + \alpha^2  \sum_{\mathbf{n}^\prime} \int_0^t d\tau \mathcal{K}^{(2)}_{\mathbf{n}, \mathbf{n}^\prime} (t - \tau) \rho_{\mathbf{n}^\prime}(\tau).  
\end{align}
\end{widetext}
Here the first two terms on the RHS are the regular HEOM, the last term is the second-order NZ correction for the influence of an additional environment, with NZ memory Kernel
\begin{align}
    \mathcal{K}^{(2)}_{\mathbf{n}, \mathbf{n}^\prime} (\tau) = \Tr_{E_0, \mathbf{n}}[\mathfrak{L}_{SE_0, \mathbf{n}}  e^{(\mathfrak{L}_{\mathbf{n}^\prime} + \mathfrak{L}_{E_0, \mathbf{n}^\prime})\tau} \mathfrak{L}_{SE_0, \mathbf{n}^\prime}  \sigma_{E_0}^\beta ].
\end{align}

We note that this involves a summation over reduced ADOs from every tier, however, the only term that mixes tiers is $e^{\mathcal{L}_\mathbf{n^\prime} \tau}$, which acts, as one may expect, to couple tiers of increasing distance as time evolves. For example, if we are to expand this term as
\begin{align}
    e^{\mathfrak{L}_\mathbf{n} \tau} = \mathbf{1}_\mathbf{n} + \mathfrak{L}_\mathbf{n} \tau + \frac{1}{2}(\mathfrak{L}_\mathbf{n} \tau)^2 + \cdots
\end{align}
we see that each term couples tiers with an additional separation of a maximum 2 (one above, if available, and one below). In Ref. \cite{fay2022} a similar approach has been exploited, and a Redfield like term is obtained making the Born-Markov approximation of the NZ kernel. 

Now, we note that $e^{\mathfrak{L}_\mathbf{n} \tau}\mathfrak{L}_{SE_0, \mathbf{n} + \mathfrak{L}_{E_0, \mathbf{n}^\prime}}  = \tilde{\mathfrak{L}}_{SE_0, \mathbf{n}} (\tau)$, where the tilde denotes an interaction picture representation, where the `system' is the uncoupled full hierarchy, and the environment is $E_0$. Then,
\begin{align}
    \mathcal{K}_{\mathbf{n}, \mathbf{n}^\prime} (\tau) &= \Tr_{E_0}[\mathfrak{L}_{SE_0, \mathbf{n}} \tilde{\mathfrak{L}}_{SE_0, \mathbf{n}^\prime} (\tau) \sigma_{E_0}^\beta ] \nonumber \\& 
    = - \Tr_{E_0}[ [H_{SE_0, \mathbf{n}}, [\tilde{H}_{SE_0, \mathbf{n}^\prime}(\tau), \sigma_{E_0}^\beta]] ] \\&
    = - \Tr_{E_0}[ [s_{E_0, \mathbf{n}} B_{E_0, \mathbf{n}} , [\tilde{s}_{E_0, \mathbf{n}^\prime}(\tau) \tilde{B}_{E_0, \mathbf{n}^\prime}(\tau), \sigma_{E_0}^\beta]] ]. \nonumber
\end{align}
Then, defining the free environmental correlation functions $C_{\mathbf{n}, \mathbf{n}^\prime}^{(E_0)}(\tau) = \Tr_{E_0}[\tilde{B}_{E_0, \mathbf{n}^\prime}(\tau) B_{E_0, \mathbf{n}^\prime}  \sigma_{E_0, \mathbf{n}^\prime}^\beta]$ we have for an interaction Hamiltonian $H_{SE, \mathbf{n}} = \alpha s_{\mathbf{n}}B_{\mathbf{n}}$ (we note that the operators are identical for each tier $\mathbf{n}$, though the label is nonetheless useful to keep track of the ADO tiers on which they act)
\begin{align}
    \mathcal{K}_{\mathbf{n}, \mathbf{n}^\prime} (\tau) \rho_{\mathbf{n}^\prime} &= s_{\mathbf{n}}s_{\mathbf{n}^\prime}(t-\tau) \rho_{\mathbf{n}^\prime} (\tau)C_{ \mathbf{n}, \mathbf{n}^\prime}^{(E_0)}(\tau - t ) \nonumber \\&
    - s_{\mathbf{n}} \rho_{\mathbf{n}^\prime} (\tau)s_{\mathbf{n}^\prime}(t-\tau)C_{ \mathbf{n}^\prime, \mathbf{n}}^{(E_0)}(t - \tau ) \\&
    - s_{\mathbf{n}^\prime}(t-\tau) \rho_{\mathbf{n}^\prime}(\tau)  s_{\mathbf{n}}C_{ \mathbf{n}, \mathbf{n}^\prime}^{(E_0)}(\tau - t ) \nonumber \\&
    + \rho_{\mathbf{n}^\prime} (\tau) s_{\mathbf{n}^\prime}(t-\tau) s_{\mathbf{n}}C_{ \mathbf{n}^\prime, \mathbf{n}}^{(E_0)}(t - \tau ) \nonumber
\end{align}
We can then take the Markovian limit, where $C_{ \mathbf{n}, \mathbf{n}^\prime}^{(E_0)}(\tau) = \delta(\tau)\delta_{\mathbf{n}, \mathbf{n}^\prime}$, and extending upper limit of the time integral to infinity, we can write the correction term $\Gamma_{\mathbf{n}, \mathbf{n}^\prime} = \int_0^\infty d\tau \mathcal{K}_{\mathbf{n}, \mathbf{n}^\prime} (\tau) \rho_{\mathbf{n}^\prime}$ as
\begin{align}
    \Gamma_{\mathbf{n}, \mathbf{n}}(t) = s_{\mathbf{n}}s_{\mathbf{n}}\rho_{\mathbf{n}} (t) &\nonumber 
    - s_{\mathbf{n}} \rho_{\mathbf{n}} (t)s_{\mathbf{n}}
    - s_{\mathbf{n}} \rho_{\mathbf{n}}(t)  s_{\mathbf{n}}\nonumber \\&
    + \rho_{\mathbf{n}} (t) s_{\mathbf{n}}s_{\mathbf{n}}
\end{align}
Now, if $E_0$ is an optical environment coupled via, say $s = \sigma_x = \sigma_+ + \sigma_-$, we can take a rotating wave approximation, and recover a GKSL type term for each tier:
\begin{align}
        \Gamma_{\mathbf{n}, \mathbf{n}}(t) = 2\sigma_- \rho_{\mathbf{n}} (t)  \sigma_+ 
    - \{\sigma_+\sigma_-,  \rho_{\mathbf{n}}(t)  \} =: \mathcal{D}_\mathbf{n}[\rho_\mathbf{n}(t)]
\end{align}
and we have the GKSL-HEOM of the form
\begin{align}\label{eq:GKSL-HEOM_fin}
   \partial_t \rho_{\mathbf{n}}(t)= (\mathcal{L}^D_{\mathbf{n}}  & + \mathcal{D}_{\mathbf{n}})\rho_{\mathbf{n}}(t) \\& + \sum_k^K ( \mathcal{L}^+_{\mathbf{n}}\rho_{\mathbf{n}_{k+
}}(t) +  \mathcal{L}^{-}_{\mathbf{n}}\rho_{\mathbf{n}_{k-}}(t))\nonumber
\end{align}
We thus see that the standard set of approximations that lead to a GKSL equation yield non-additive dynamics with the HEOM - that is, the resulting dissipators act on every tier of the ADOs.

We further note that in the most general form of the NZ-HEOM, Eq. \eqref{eq:NZ-HEOM_general}, the memory kernel may act to couple tiers of the hierarchy, and even ADOs within the same tier not otherwise coupled via standard HEOM.

\section{Nakajima-Zwanzig Equation}\label{app:NZ}

Here, for completeness, we derive the NZ quantum master equation without the presence of the non-perturbative environment described by HEOM. We will largely follow references \cite{BRE02, gardiner00, brian2021}, and refer the reader to \cite{brian2021} for a more detailed review. We begin by writing the equations of motion for the relevant and irrelevant parts of the total density matrix by applying the projection operators directly to the LvN equation,
\begin{subequations}\label{eq:NZ_parts}
\begin{eqnarray}
  \frac{d}{d t} \mathcal{P} \rho(t) = \mathcal{PLP}\rho(t) +  \mathcal{PLQ}\rho(t)   \\
  \frac{d}{d t} \mathcal{Q} \rho(t) = \mathcal{QLP}\rho(t)   + \mathcal{QLQ}\rho(t) 
\end{eqnarray}
\end{subequations}
where we have used $\mathcal{P + Q} = \mathbb{1}$, and that both projection operators commute with the derivative. We than define the Laplace transform $\tilde{f}(s) = \mathfrak{L}\{f(t)\}(s) = \int_0^\infty st e^{-st} f(t)$, which has the useful properties,
\begin{align}\label{eq:Laplace_props}
    & \mathfrak{L}\left\{\frac{df(t)}{dt}\right\}(s) = \int_0^\infty e^{-st}\frac{df(t)}{dt}dt = s\tilde{f}(s) - f(0) \nonumber \\ &
    \mathfrak{L}\{g(t) \star f(t) \} = \tilde{g}(s) \tilde{f}(s),
\end{align}
where we have defined the convolution $g(t) \star f(t) = \int_0^\infty d\tau g(t - \tau) f(\tau)$. Using the first of these properties we can take the Laplace transforms of Eq. \eqref{eq:NZ_parts}, obtaining 
\begin{subequations}\label{eq:NZ_parts_laplace}
    \begin{eqnarray}
      s\mathcal{P} \tilde{\rho}(s) - \mathcal{P} \rho(0) = \mathcal{PLP}\tilde{\rho}(s) + \mathcal{PLQ}\tilde{\rho}(s) \label{eq:NZ_parts_laplace_a} \\
      s\mathcal{Q} \tilde{\rho}(s) - \mathcal{Q} \rho(0)  = \mathcal{QLP}\tilde{\rho}(s) + \mathcal{QLQ}\tilde{\rho}(s) \label{eq:NZ_parts_laplace_b}.
    \end{eqnarray}
\end{subequations}
Rearranging Eq. \eqref{eq:NZ_parts_laplace_b} we obtain
\begin{align}
    \mathcal{Q}\tilde{\rho}(s) = \frac{1}{s - \mathcal{QL}} \mathcal{Q} \left(\rho(0) + \mathcal{LP} \tilde{\rho}(s)  \right)
\end{align}
which we can substitute into Eq. \eqref{eq:NZ_parts_laplace_a} to find
\begin{align}
      s\mathcal{P} \tilde{\rho}(s) - \mathcal{P} &\rho(0) = \mathcal{PLP}\tilde{\rho}(s) \nonumber \\&
      + \mathcal{PL}\frac{1}{s - \mathcal{QL}} \mathcal{Q} \left(\rho(0) + \mathcal{LP} \tilde{\rho}(s)  \right).
\end{align}
We then take the inverse Laplace transform of this expression, which for the LHS and first term of the RHS is trivial (as they are themselves the result of a Laplace transform above), and for the second term of the RHS requires use of the convolution property in Eq. \eqref{eq:Laplace_props}. We thus obtain
\begin{align}\label{eq:NZ_general}
    \frac{d}{dt} \mathcal{P} \rho(t)& = \mathcal{PLP}\rho(t) + \mathcal{PL} e^{\mathcal{QL}t} \mathcal{Q} \rho(0) \nonumber \\& 
    \qquad + \int_0^t d\tau \mathcal{PL} e^{\mathcal{QL}(t - \tau)} \mathcal{QLP} \rho(\tau) \\&
     =   \mathcal{PLP}\rho(t) + \mathcal{I}(t) + \int_0^t d \tau \mathcal{K}(t - \tau) \mathcal{P}\rho(\tau) \nonumber.
\end{align}
Where we have defined the inhomogeneous term $\mathcal{I}(t)$, which vanishes is $\mathcal{P} \rho(0) = \rho(0)$, and the memory kernel $\mathcal{K}(t)$. 

We note that up to this point we have made no reference to the form of the projection operators $\mathcal{P}, \, \mathcal{Q}$, and thus the result is equally applicable to the HEOM case. First, however, we simplify the result using the more standard projection operators defined by $\mathcal{P} \rho = \Tr_E[\rho] \otimes \rho_E$, and $\mathcal{Q} = \mathbf{1} - \mathcal{P}$. In this case, we can further simplify the problem, 
\begin{align}\label{eq:NZ}
    \frac{d}{dt} \rho_S(t) =  \mathcal{L}_S \rho_S(t) 
    + \int_0^t d\tau \mathcal{K}_S(t - \tau) \rho_S(\tau), 
\end{align}
where $ \mathcal{K}_S(t) := \Tr_E[\mathcal{PL} e^{\mathcal{QL}t} \mathcal{QLP}]$.

\subsection{Weak coupling limit}

We now wish to take the weak coupling and Markovian limits of Eq. \eqref{eq:NZ}, which is straightforwardly obtained by once again looking at the Laplace transformed equations for the relevant and irrelevant subspaces in Eq. \eqref{eq:NZ_parts_laplace}. We then redefine $\mathcal{L} = \mathcal{L}_S + \mathcal{L}_E + \alpha \mathcal{L}_I$, such that the parameter $\alpha$ determines the coupling strength to the environment. We can then use that $\mathcal{PL}_S = \mathcal{L}_S \mathcal{P}$, $\mathcal{PL}_E = \mathcal{L}_E \mathcal{P} = 0$, and $\mathcal{P}\mathcal{L}_I \mathcal{P} = 0$ to obtain
\begin{subequations}\label{eq:projection_permutations}
    \begin{eqnarray}
          &\mathcal{PLP} = \mathcal{P\mathcal{L}_S P} \\&
          \mathcal{PLQ} = \alpha \mathcal{PL}_I \mathcal{Q}\\&
          \mathcal{QLP} = \alpha \mathcal{QL}_I \mathcal{P}\\&
        \mathcal{QLQ} = \mathcal{L}_S\mathcal{Q} + \mathcal{L}_E\mathcal{Q} + \alpha\mathcal{QL}_I\mathcal{Q}.
    \end{eqnarray}
\end{subequations}
We then substitute these expressions into Eq. \eqref{eq:NZ_parts_laplace_b} and solve for $\mathcal{Q}\rho$ as before, obtaining
\begin{align}
    \mathcal{Q} \tilde{\rho}(s)  = \frac{1}{s -  (\mathcal{L}_S + \mathcal{L}_E + \alpha\mathcal{QL}_I) } \alpha \mathcal{L}_I \mathcal{P}\tilde{\rho}(s) 
\end{align}
where we have once again ignored the term $\propto  \mathcal{Q} \rho(0)$, as this vanished for factorising initial conditions. This can be substituted once more into Eq. \eqref{eq:NZ_parts_laplace_a} alongside Eqs. \eqref{eq:projection_permutations} to find
\begin{align}
          s\mathcal{P} \tilde{\rho}(s) - &\mathcal{P} \rho(0)=  \mathcal{P}\mathcal{L}_S \mathcal{P}\tilde{\rho}(s) \\  & 
          + \alpha^2 \mathcal{PL}_I \frac{1}{s -  (\mathcal{L}_S + \mathcal{L}_E + \alpha\mathcal{QL}_I) }\mathcal{QL}_I \mathcal{P}\tilde{\rho}(s) \nonumber.
\end{align}
we then take the lowest order in $\alpha$ in each the homogeneous and inhomogeneous terms,  \cite{gardiner00}
\begin{align}
          s\mathcal{P} \tilde{\rho}(s) &= \mathcal{P} \rho(0) + \mathcal{P}\mathcal{L}_S \mathcal{P}\tilde{\rho}(s) \\  & 
          + \alpha^2 \mathcal{PL}_I \frac{1}{s -  (\mathcal{L}_S + \mathcal{L}_E )}\mathcal{L}_I \mathcal{P}\tilde{\rho}(s) \nonumber.
\end{align}
Finally, once again exploiting the convolution property of the Laplace transform we may take its inverse and trace over the environment to find
\begin{align}
    \frac{d}{dt} \rho_S(t) & = \mathcal{L}_S \rho_S(t) \nonumber \\ & + \alpha^2\int_0^t d\tau \Tr_E[\mathcal{PL}_I  e^{(\mathcal{L}_S + \mathcal{L}_E)\tau} \mathcal{L}_I \mathcal{P}\rho(t - \tau)]  \nonumber \\&
    = -i [H_S, \rho_S(t)] \\& - \alpha^2 \int_0^t d\tau Tr_E[ [H_I(t), [H_I(t - \tau), \rho_S(t - \tau) \otimes \rho_B]]] ,\nonumber 
\end{align}
where $H_I(\tau) = e^{(\mathcal{L}_S + \mathcal{L}_E) \tau}H_I$ is the interaction Hamiltonian in the interaction picture. We thus recover the familiar Bloch-Redfield equation from which we may straightforwardly obtain the GKSL master equation \cite{BRE02}.

\end{document}